\documentclass[a4paper,fleqn,usenatbib]{mnras}

\usepackage{mathptmx}
\usepackage{color}

\usepackage[T1]{fontenc}
\usepackage{ae,aecompl}

\usepackage{graphicx}	
\graphicspath{{./plots/}}

\usepackage{amsmath}	
\usepackage{amssymb}	
\usepackage{epstopdf}
\usepackage{changes}

\usepackage{hyperref} 
\hypersetup{
    colorlinks=true,       
    linkcolor=blue,          
    citecolor=blue,        
    urlcolor=blue           
}

\newcommand{\msun}{\text{M}_{\odot}}
\newcommand{\SFR}{\text{M}_{\odot} \text{yr}^{-1}}
\newcommand{\Zsol}{\text{Z}_{\odot}}
\newcommand{\Ha}{\text{H}\alpha}
\usepackage{color}

\title[Dwarf Galaxy Evolution - The impact of the IMF]
{Why should Models of Dwarf Galaxy Evolution care about the Initial Mass Function at low Star-formation Rates?}

\author[Steyrleithner \& Hensler]{
P. Steyrleithner\thanks{E-mail: steyrleitp21@univie.ac.at},
G. Hensler
\\
Department for Astrophysics, University of Vienna, T\"urkenschanzstrasse 17, A-1180 Vienna, Austria\\
}

\date{Accepted XXX. Received YYY; in original form ZZZ}

\pubyear{2015}

\begin{document}
\label{firstpage}
\pagerange{\pageref{firstpage}--\pageref{lastpage}}
\maketitle

\begin{abstract}
When star clusters are formed at low star-formation rates (SFRs), their stellar initial mass function (IMF) can hardly be filled continuously with stars at each mass. This lack holds for massive stars and is observationally verified by the correlation between star-cluster mass and its most massive cluster star. 
Since galaxy evolution is strongly affected by massive stars, numerical models should account for this lack. Because a filled IMF is mostly applied even when only fractions of massive stars form, here we investigate by 3D chemo-dynamical simulations of isolated dwarf galaxies how deviations from a standard IMF in star clusters affect the evolution.
We compare two different IMF recipes, a filled IMF with one truncated at a maximum mass at which a single complete star forms. 
Attention is given to energetic and chemical feedback by massive stars.
Since their energy release is mass dependent but steeper than the negative IMF slope, the energetic feedback retains a positive mass dependence, so that a filled IMF regulates SF stronger than truncated IMFs, though only stellar number fractions exist. 
The higher SFR of the truncated IMF in the simulation leads to more supernovae II (SNeII), driving galactic winds. Whether this results from the model-inherent larger SFR is questioned and therefore analytically explored. This shows the expected result for Lyman continuum, but that the total SNII energy release is equal for both IMF modes, while the power is smaller for the truncated IMF. 
Reasonably, the different IMFs leave fingerprints in abundance ratios of massive-to-intermediate-mass star elements.
\end{abstract}




\section{Introduction}

Stars are born in galaxies obeying a particular mass distribution, the so-called initial mass function (IMF). Originally defined by \citet{Salpeter55} as the number of stars $N(m)$ in a mass bin dm from local star counts, it was believed for a long time that this IMF should be universal. 
Individual observations and comprehensive surveys with advanced techniques over the last two decades, provide deeper insights into the IMF of star clusters, but also raise doubts on the universality of the IMF.

The IMF is one of the most important distribution function in astrophysics, because it determines the mass-to-light ratio $M/L$ and the baryonic fraction of the dynamical mass of galaxies as well as the transformation efficiency of gas to stars due to the mass-dependent stellar feedback. 
It is widely believed that the majority of stars is formed in associations, e.g. \cite{LadaLada03}. In molecular clouds each star/binary is formed from the individual denser cores (clumps) which consist of sizes and mass scales comparable with those of individual star clusters. The core-mass function (CMF) is already close to the IMF \citep{LAL15}. How the CMF develops from the divergent cloud mass function and how it transforms to a universal IMF is still a field of active debate and rapid progress \citep{Krum19,And19,Pel21}.
 
As part of the IMF, massive stars have the shortest lives and release the most energy, mass, and heavy elements per stellar mass to their environment. Vice versa, the long-living, less active low-mass stars accumulate the stellar mass budget. 
By this, the IMF governs the evolution of galaxies and star clusters and determines the mass-to-light ratio $M/L$ of galaxies. 
For stars more massive than the sun, the IMF can be approximated by a single power-law of $-2.35$ \citep{Salpeter55}. 
More refined analyses accounting also for the low-mass range \citep{Kroupa02} have unveiled a multi-power-law function as more appropriate with a lower and more evenly inclining slope at mass ranges below $0.1\,\msun$. Today the IMF by \citet{Kroupa01} with a three-part power-law is widely used and considered equal to that of \citet{Cha03}. 

The mass distribution of embedded star clusters $N(M_{ecl})$ can also be described by a single power-law with increasing number of low-mass clusters (see discussion in \citet{Elm06}). 
In addition, there are two further correlations which are deduced from observations and are of relevance for the strategy of stellar mass distributions:
i) Reasonably, the maximum young cluster mass $M_{cl,max}$ in a galaxy depends on the galactic star-formation rate (SFR) as found by \cite{Lar02, WK05,JSD17} for a correlation of $M_{cl,max}$ with the SFR column density $\Sigma_{SFR}$. 
ii) Compiling the most massive star $m_{max}$ of star clusters with their masses $M_{cl}$ from the literature, \cite{WKB10} find a striking correlation comparable to those derived analytically by \cite{Elmegreen2000} and modeled by \cite{BBV03}. However, \cite{ACC13,ACC14,Weisz15,DSH17} find no evidence of a $m_{max}$ dependence on the cluster mass. To study the existence of $m_{max}$ \cite{PH2014} perform 25 million simulations of random stellar mass distributions and find that an uppermost mass range exists instead of a single $m_{max}$ and also analytically derive a relation of $m_{max} \propto M_{cl}^{1/1.35}$.

\citet{WK06} performed Monte Carlo simulations to establish this $m_{max}-M_{cl}$ correlation observationally. They randomly take clusters from an embedded cluster mass function and fill them with stars from an IMF either with or without an upper mass limit, thereby reaching single-star or randomly sampled or sorted clusters. They come to the conclusion that an IMF populated with stars randomly and without a mass limit do not match the observations. The scatter in $m_{max}$ with the cluster mass can thereby be understood to emerge from observational uncertainties only \citep{WKP13}.  

As a consequence of correlation ii), a lack of massive stars has to exist in low-mass clusters, called top-light IMF, and equal to a dominance of low-mass stars, called bottom-heavy IMF, if one focusses on the low-mass part. This fact is observationally manifested through various indicators, e.g. by stellar population analysis of galaxies with different masses, low-surface brightness galaxies (LSBs) \citep{Lee04}, early-type galaxies (ETGs) \citep{CvD12}, and massive cluster galaxies \citep{Lou21}, or by the spectral signatures as there are $\Ha$ vs. UV \citep{Cal10,Lee09}, and furthermore, by chemical abundance analyses \citep{Tsu11,McW13}. Moreover, ETG studies also reveal possible correlations of the IMF slope with velocity dispersion \citep{Fer13,MaNa15} and metallicity \citep{MaNa15} in the sense of a steeper IMF slope with larger velocity dispersion and metallicity, respectively.

Dwarf galaxies (DGs) are the most numerous type of galaxies in the universe. They have low surface brightness, low gravitational potential wells and relatively low SFRs. This latter property questions whether the IMF can be filled up to the most massive stars or is truncated at a low but massive limit according to ii). In fact, \citet{Lee04} find a much steeper power-law exponent than Salpeter or Kroupa and favored a bottom-heavy IMF. Due to their low gravitational potential wells, DGs react more vigorously to external and internal processes such as feedback by massive stars and are therefore ideal objects to study galaxy evolution. 

\citet{Lee09} compare SFRs derived from the $\Ha$ flux, a tracer for the most massive O-stars $(m_\ast \ge 17\,\msun)$, with that from the FUV flux, a tracer for O-to-early-B stars $(m_\ast \ge 3\,\msun)$ of a sample of $\sim 300$ star-forming galaxies within 11 Mpc distance from the Milky Way. They find that $\Ha$-to-FUV flux ratios are up to one order of magnitude lower than expected for SFRs below $10^{-2} \, \msun\, \text{yr}^{-1}$. They conclude that an IMF which is deficient of massive stars in DGs and LSBs is consistent with their data because the $\Ha$ luminosity flattens at very low SFRs \citep{PWK07}. This $\Ha$-to-FUV flux divergence at low SFR is also documented by \citet{Meu09} as is the decreasing ratio of stellar ionizing to non-ionizing flux with smaller galaxy luminosity and mass, as well as a lower SFR \citep{Meu09,Bos09}. \citet{Cal10} find that a tendency towards a decreasing $\Ha$ luminosity-to-cluster mass $M_{cl}$ exists, but not as strong as expected from ii).
Although by a simple understanding, the $\Ha$-to-UV discrepancy of the derived SFRs can be explained by the lack of massive stars caused by a truncated and/or a steeper IMF, additional explanations are elaborated. To solve this discrepancy, \cite{FGS11} study the stochastic sampling for a universal IMF at low SFRs of DGs and vary the fractions of cluster-born vs. field stars. Another strategy is proposed by \cite{Weisz12}, applying short-term (+10 Myr) SFR episodes timely separated by one to a few hundred Myr, which matches the observations much better than a variable IMF alone. A further successful exploration is performed by \cite{Eldridge12} for a purely stochastic sampling of the IMF with the inclusion of binary-star formation.
Interestingly, \citet{Roy09} find from UV data of extremely faint DGs taken with the GALEX satellite, no threshold of the SFR with the gas column density $\Sigma_{HI}$ as it is suggested by \citet{Ken98}. $\Sigma_{SFR}$ reaches down to $10^{-6}\,\msun\,\text{yr}^{-1}\,\text{kpc}^{-2}$ and is only limited by the GALEX sensitivity. These results support the $m_{max}-M_{cl},\,M_{cl,max}-\text{SFR}$ correlations explained by the hierarchical approach of SF \citep{EE14}. 

In addition to that, \citet{WKP13} argue that fainter galaxies have steeper IMF slopes than brighter (more massive) ones. This agrees with observational hints that massive galaxies with large SFRs are sites of a top-heavy IMF \citep{GAMA11,KWP13,NLO05}. \cite{BLF05} also find from semi-analytical models using submillimeter observations that the observed counts from early submillimeter galaxies (SMG) samples can only be matched by adopting a very top-heavy IMF. \cite{NLO05} stated that only extreme top-heavy IMFs can explain the properties of elliptical galaxies and \cite{WKP13} suggest that "an early strong starbursting stage with a top-heavy IMF" must have existed for ETGs, followed by a second epoch with (probably) a bottom-heavy IMF. This argument for a coexistence of both possibilities is also pointed out in the review of \cite{Smith20}.

If the SF is concentrated to a single region, two extreme cases of SF are observed. Star clusters with very low SFRs can e.g. be formed in very dilute gas environments like e.g. in tidal tails of interacting galaxies, e.g. \citet{LeeW16,LeeW18}, and in ram-pressure stripped gas clouds of cluster galaxies, e.g. \citet{Bos18}. A rough estimate already prohibits an IMF from being completely populated \citep{PHR14}. The other extreme is so-called super star clusters produced in starburst DGs, which seem to violate the $M_{cl,max}$-SFR relation. Moreover, the deterministic vs. stochastic massive star formation is under debate observationally \citep{ACC14} and theoretically (see e.g. reviews by \citet{KroupaRev14} and \citet{Krumholz14}). That the stochastic population of e.g. $100$ clusters of $1000\,\msun$ must produce the same numbers and masses of massive stars as a single $10^5\, \msun$ cluster and represent a fully sampled IMF is in contrast to the $m_{max}-M_{cl}$ and $M_{cl,max}-\text{SFR}$ correlations. These require a smaller total number of massive stars in numerous clusters than the single cluster, which must lead to differences in the stellar feedback, in terms of energetic as well as chemical yields. 

On the other hand, the validity of $\Ha$ as SFR equivalent must be also questioned, because stellar feedback can cause the $\Ha$ flux to both underestimate \citep{ACC13} and overestimate \citep{MRH15} the SFR.
\cite{ACC14} demonstrate strikingly that a limitation of the uppermost stellar mass leads to a reduced $\Ha$ emission from a star cluster as a result of lacking Lyman continuum photons. In DGs superbubbles from accumulated type II supernovae (SNeII) can open holes in the interstellar medium (ISM) around star clusters and allow the escape of Lyman continuum ($Ly_c$) photons. This leakage has to depend on the shape of the DG disk \citep{RH13}. Analysing these models \citet{MRH15} find an escape fraction of about $40\,\%$, but also that additional $Ly_c$ photons from cooling hot superbubbles raise the $\Ha$ emission by a factor of about $1.5$. 

Numerous numerical simulations of DGs have been performed over the last two decades studying the influence of SF on the DG evolution, but also the various internal/external processes that determine the formation and evolution of isolated DGs. \cite{VDD08} identify the role of angular momentum as a second parameter \citep{Schroy11}, and according to this, the minimum disk column density for SF \citep{VRH12} and the interplay between mass and geometry as important effects of galactic wind and the distribution of heavy elements \citep{RH13}. Further focus was set to the formation and evolution of particular morphological DG types with regard to SF and chemical abundances such as e.g. tidal-tail DGs (TDGs) \citep{PHR14,PRH15,Baum19}, ram-pressure stripped DGs \citep{Steyr20} and also of dwarf ellipticals (dEs) \citep{HTG04}, applying advanced chemo-dynamical multi-phase numerical codes. 
These simulations include self-consistently radiative gas cooling, SF stellar feedback by radiation, winds, type Ia and II SNe. Those yield temporarily very low SFRs. Since it is computationally too demanding to resolve single stars in a galactic simulation, formed star clusters have to be treated as single stellar populations (SSPs) containing stars of all masses within an IMF range.

To determine the correct feedback from stars, depending on their mass and lifetime, the IMF must be implemented properly. Since the IMF is thought to be invariant through large ranges of conditions \citep{Kroupa01}, this assumption is used to describe individual clusters numerically. 
For simplicity, most simulations take the universal but fully populated IMF into account, whereas, especially for low-mass DGs with their low SFRs, the formed star-cluster mass is too small to fill the IMF. Filling means that each mass bin, independent of the binning width, up to $m_{max}$ contains at least one star of that mass. Therefore in this case, two options are plausible to populate the IMF: either it can be truncated at an uppermost mass $m_{max}$ that holds at least one integer star, or the IMF is filled stochastically. 
\cite{Elme09} shows that even the latter mode mostly fills the IMF of low-mass star clusters up to a characteristic upper mass that stays, however, below the possible maximum. How a filled vs. truncated IMF affects the evolution of TDGs is already demonstrated by us \citep{PHR14}. 
For a truncated IMF, these models experience a stronger SF and thus form more stars as a consequence of lower self-regulating energy. The energetics of both IMF modes are also compared on the basis of stellar lifetime, i.e. as feedback power, but without a quantitative analysis. 

\citet{Bekki13} performs 3D hydrodynamical simulations of galaxy evolution with a universal and a non-universal IMF, respectively, and finds that the SFRs in the non-universal IMF model are lower than for the universal IMF. He also show that the IMF slopes can vary in different galaxies, but also in different local SF regions within the same galaxy. 
Having spotted this problem, \citet{App20} model DGs, applying a cosmological Smoothed-particle Hydrodynamics (SPH) code with feedback from massive stars and gas physics, and implement a self-consistent stochastically populated IMF. In comparison, they find that a continuous IMF yields a higher stellar mass, which is caused by the lower feedback of discrete supernova explosions due to overcooling and which result in a higher SFR. Unexspectedly, they do not find significant variations of chemical abundances. 
IMF variations at the massive-star range must become discernible in particular ratios of elements enriched by intermediate-mass stars (IMS) vs. massive stars. Most obviously, the abundance ratio of C produced by IMS to O released by massive stars should be sensitive to an uppermost mass cut-off or a top-down of the IMF. \citet{TsuB11} analyze C/O  ratios of Damped Lyman Alpha systems (DLAs) and conclude from the observed high values that these indicate a nucleosynthesis production dominated by stars less massive than $20-25\,\msun$.

Since the energetic feedback of massive stars by both, SNeII, stellar radiation and winds, plays a substantial role in DG evolution, the mass range of massive stars is a sensitive ingredient for galactic energetics and chemistry. The feedback of massive stars becomes even more pronounced in the shallower gravitational potential of DGs which facilitates galactic winds. 

Equally sensitive impacts of different stellar masses come from their chemical yields, because those are clearly dependent on the stellar progenitor mass, so that abundance ratios of characteristic elements attributable to intermediate vs. massive stars bear the key to deviations of the IMF.

This paper aims at quantifying the effect of the numerical simplification, assuming a filled IMF while reaching only low SFRs, so that only stellar mass fractions are theoretically populating the massive-star range. Moreover, we aim at sensitizing galaxy modelers to the evolutionary impact of lacking massive stars in low-mass star clusters. In addition, we wish to understand and quantify the (former) findings in TDG models \citep{PHR14} as to why the stellar feedback to the local star-formation process is stronger for a truncated IMF and whether its SNII energy release is only caused by the larger SFR or more by the integer numbers of SNeII explosions. 
The paper is therefore structured as followed: Sec. \ref{sec:simulation} gives an overview of the code used, the initial conditions and the implemented processes such as SF, cooling and feedback. Sec. \ref{sec:results} describes the results which are discussed in sec. \ref{sec:discussion}, with the focus on an analytical understanding of the numerical reuslts. Finally, sec. \ref{sec:conclusion} draws conclusions from our models.

\section{Simulations}\label{sec:simulation}

\subsection{The Code}
In this work we use our simulation code cdFLASH, which is an extentions of the FLASH code version $3.3$ \citep{Fryxell00} by several chemo-dynamical ingredients such as, amongst others (see the following subsections!), gas-dependent star-formation recipe, stellar-mass dependent release and trace of chemical element abundances, and their effect on gas cooling. These code extensions are successfully applied to various astrodynamical aspects like TDG evolution \citep{PHR14,PRH15,Baum19} and ram-pressure stripping \citep{Steyr20}. A repetition of the FLASH code ingredients and specifications is avoided here and can be inspected in \citep{Steyr20}. 

\subsection{Initial Conditions}\label{subsec:IC}

\begin{figure}
  \includegraphics[width=\columnwidth]{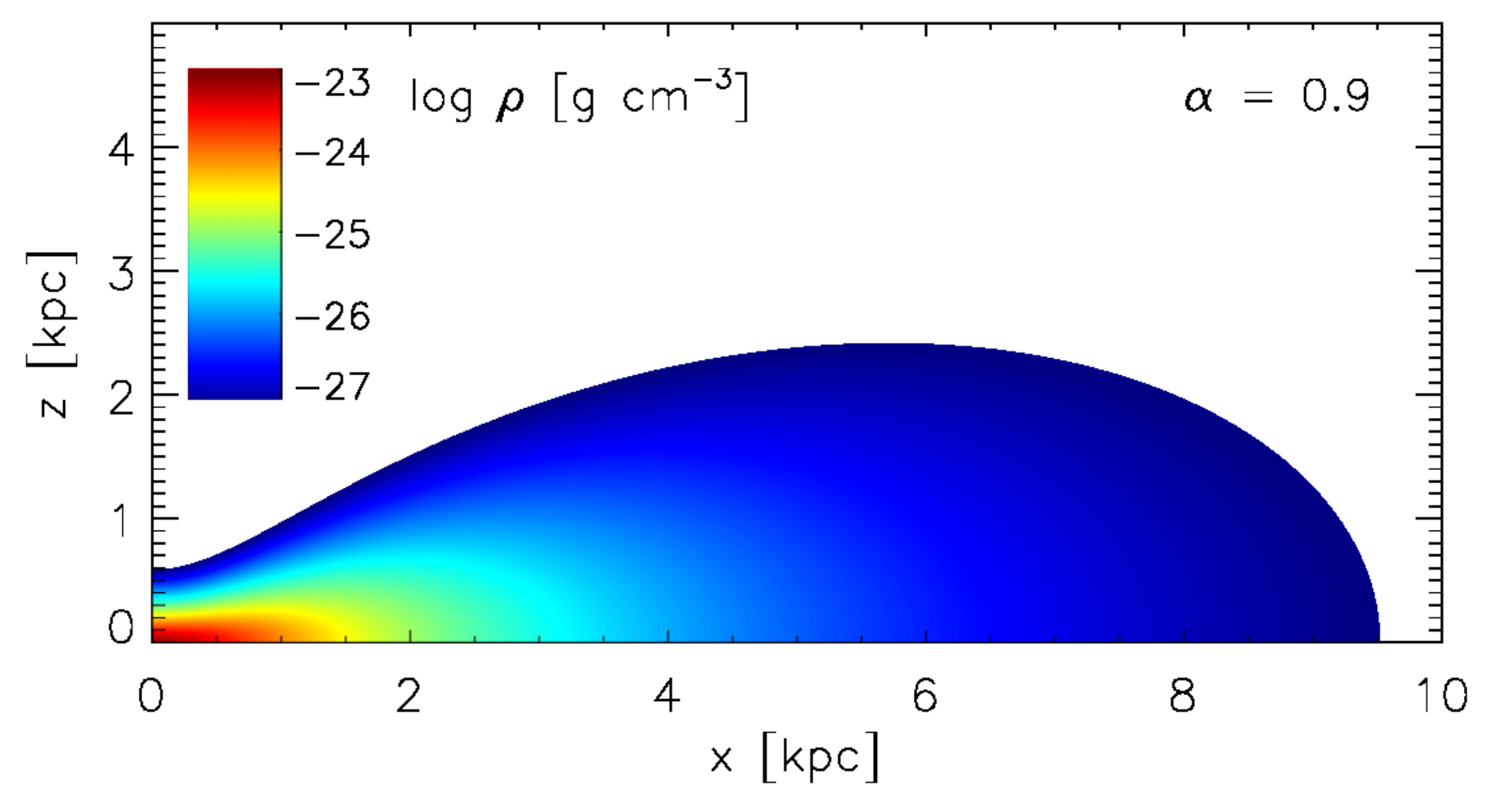} 
	\caption{Initial gas density in an edge-on cut through the galaxy center. The colour bar gives the logarithmic scale of the volume densities. }
	\label{fig:initial}
\end{figure}

Our models are aimed at starting with a purely gaseous disk, embedded in an existing stationary DM halo and are the same as in \cite{Steyr20} (see fig.\ref{fig:initial}). Since the energetics of SF and gas dynamics are determined by the gas structure itself and the total gravitational potential, we include for simplicity an old existing stellar population into the DM mass. For a clear determination of the SF effects by the massive stars of the IMF modes, the mass and energy contribution by the old population is assumed to be terminated. 
Although we are aware that this is not truly realistic cosmological scenario, having also neglecting a temporal mass growth of such low-mass systems, for our purposes it is sufficient to focus on the newly formed stars and their effects. The models are executed in a 3D box of $\pm 12\,\text{kpc}$ grid size and a six-level mesh refinement so that the highest resolution reaches $50\,\text{pc}$, i.e. less than $1/3$ of the disk scaleheight. This spatial resolution is also justified by the observed  scales of star-forming molecular clouds and the range of separations of young star clusters.

For the initial conditions of a stable disk the code described by \cite{VRH12} is used to calculate an equilibrium configuration of a rotating gas disk, where the steady-state momentum equation for the gas component in a gravitational potential due to gas and DM is solved, as described in detail in \citet[see sec. 2.2]{Steyr20}. Although the initial mass setup is analytically an equilibrium state, we relax the model numerically for another $230\,\text{Myr}$ before the physical processes as e.g. SF, cooling, etc. are switched on.

Most simulations of isolated galaxies neglect self-gravity in building the initial configuration, so our approach is an improvement compared to commonly used methods \citep{VRH12,VRH15}. For the DM halo a spherical isothermal density distribution is assumed.

We simulate two DGs with identical initial conditions: In one simulation all mass bins of the IMF are filled, according to the Kroupa IMF, until an upper mass (that does not depend on the cluster mass) of $m_{max} = 120\,\msun$, allowing the formation of fractions of massive stars. In the other simulation the IMF is truncated at an uppermost mass, at which the next bin would contain less than one full star. 

For a given halo mass of $M_{DM} = 10^{10}\,\msun$ and a spin parameter of $\alpha = 0.9$ the resulting gas mass is $M_{g} = 1.4\times10^{8}\,\msun$ with a maximum rotation velocity of $v_{rot} = 30\,\text{km}\,\text{s}^{-1}$. The radius $R_{gal}$ of the DG is defined to be the radial distance at which the gas density falls below $10^{-27}$ g cm$^{-3}$ and amounts to $R_{gal}$ = 9.5 kpc. 
For regions outside of $R_{gal}$, the gas density is set to $10^{-30}$ g cm$^{-3}$ and zhe temperature to $T = 10^6$ K to be in pressure equilibrium with the galaxy. 
The DM mass within $R_{gal}$ amounts to $M_{DM,in} = 8.4\times10^8\,\msun$.

The choice of such a low-mass DG is also made in order to get low SFR and stellar cluster masses that will not fill the stellar initial mass function. With higher gas masses the SFR increases, which will result in more massive cluster masses that fill the IMF. The initial metallicity in our simulations is based on $Z = 0.32\,\Zsol$.

\subsection{Cooling}

For the radiative gas cooling, two cooling functions are used and described in detail in \cite[see sec. 3.1]{Steyr20}. Since the stellar mass loss is also specified for the chemical elements, the cooling function above $10^4\,\text{K}$ by \citet{BH89} assumes collisional ionization equilibrium and depends on the ten most abundant elements (H, He, C, N, O, Ne, Mg, Si, S, Fe) and transits smoothly to bremsstrahlung above several $10^7\,\text{K}$. 
Below $10^4\,\text{K}$, a combination of the cooling functions by \citet{DM72} \& \citet{SKK09} is applied, for which the elements C, N, O, Si, S, Ne and Fe and the electron collisions for these elements are taken into account, and which are given by a series of equations \citep[eq. 4 - 12,][]{SKK09}. The total cooling function is the sum of the cooling of these elements.

\subsection{Star Formation}

For the SF the fully analyticly derived SFR formula of self-regulation by \cite{KTH95} is applied:
	\begin{equation} \label{eq:SBF}
		\psi(\rho,T) = C_n\, \rho^n\, f(T)
	\end{equation}
with a power-law dependence on the gas density $\rho$ and a temperature T efficiency function with an efficiency factor $C_n$. $f(T)$ is an exponential function of 
	\begin{equation} \label{eq:SBF_eff}
		f(T) = e^{-T/T_s}
	\end{equation}
that accounts for the maximum star formation at low temperatures, but a smooth transition to almost $0$ close to ionized gas. $T_s$ is therefore set to $1000\,\text{K}$. For the optimal value \cite{KTH95} found $n=2$ in agreement with the prescription by \cite{Larson88}. $C_2$ then amounts to $2.575\times 10^8\,\text{cm}^3\,\text{g}^{-1}\,\text{s}^{-1}$ so that $\psi$ of eq. \ref{eq:SBF} gets the meaning of a SFR density and is called "stellar birth function" (SBF). Since a fraction of the gas is already ionized by the stellar radiation of massive stars around star particles, the total amount of gas available for star formation has to be reduced by these HII regions and heated according to stellar feedback (see subsec. \ref{subsec:SNIIfb}) so that the SFR in a closed cell would decline from timestep to timestep.

Each SF event should produce a stellar particle that represents a single stellar population in simulations. Since eq. \ref{eq:SBF} allows SF even at very low rates leading to insignificantly low-mass star clusters and artificially numerous stellar particles, a stellar particle is generated only when a certain threshold of SFR density
	\begin{equation}\label{eq:SBF_thres}
		\psi_{thres} = \frac{ \theta_{sf} \, M_{cl,min} }{ \tau_{cl} \, V_{GMC} }
	\end{equation}
is exceeded. 
$\psi_{thres}$ combines $\theta_{sf}$ as a dimensionless factor and is set to 100, a minimum cluster mass set to $M_{cl,min}=100\,\msun$ and the cluster-formation timescale $\tau_{cl}$ of $1\,\text{Myr}$ as used in \citet{PRH15, Baum19, Steyr20} to be a reasonable timescale for the pre-main sequence formation of massive stars. $V_{GMC}$ as the volume of a giant molecular cloud (GMC) and the density are derived from the \cite{Larson88} relations. Taking e.g. a radius $R_{GMC}=200\,\text{pc}$ into account with a resulting volume of $V_{GMC}=3.35\times 10^7\,\text{pc}^3$ $\psi_{thres}$ results to  $2.984\times 10^{-4}\,\msun\,\text{Myr}^{-1}\,\text{pc}^{-3}$. While for a smaller $V_{GMC}$ $\psi_{thres}$ and, by this, also the SFR decreases, the minimum spatial resolution of 50 pc (subsect. \ref{subsec:IC}) facilitates cluster formation within the same GMC but separated by neighbouring grid cell and according with observed separations of young star clusters. 
\\
For the formation time of a star cluster we make the following Ansatz: The free-fall time for a dense core is of the order of $10^4-10^5\,\text{yr}$. As we know, this would lead to a SFR in our Milky Way molecular clouds of $\sim 100\,\msun\,\text{yr}^{-1}$. This is almost 2 orders of magnitude larger than the actual SFR, so that the formation time for stars and thus also for a cluster should be of the order of $1\,\text{Myr}$. As an order of magnitude this also accounts for the gas disruption by SNeII of the Myr-living massive stars.
\\
The short cluster-formation time $\tau_{cl}$ also ensures a quick start of the stellar feedback, which is important for a self-regulated SF. If the $\tau_{cl}$ is larger than $1\,\text{Myr}$, then the stellar particles still accrete mass from surrounding cells, which also fulfils the SF criteria, although massive stars should have already started to act by their feedback. If the SF criteria are fulfilled, stars can form, but subsequent SF will be affected by stellar feedback from previous generations.
\\
Our SF algorithm can be explained as follows: 
If the SF criteria $\psi > \psi_{thres}$ is fulfilled, the newly formed stellar particle will have the mass
	\begin{equation}\label{eq:particlaMass}
		dm = \psi \cdot dt \cdot V_{cell}\, ,
	\end{equation}
where $dt$ is the numerical time step and $V_{cell}$ the volume of the computational cell in which the SF occurs.
During the cluster-formation time of $\tau_{cl}$ the stellar particle acts as a sink particle. Each timestep the particle will accumulate mass according to eq. \ref{eq:particlaMass}, if the SF criteria is fulfilled in the corresponding cell. When the particle age exceeds the cluster-formation time $\tau_{cl}$, the sink particle will be closed for further mass accreation and start its life as a SSP with stellar feedback (see subsec. \ref{sect:feedback}). The sum of the accumulated mass of this particle is then the cluster mass $M_{cl}$.

\subsection{Uncertainties of Parameter Choices}
As in all numerical treatments of subgrid physics of star-formation parametrization a reasonable discussion of the arbitrarily chosen parameters with respect to the effects on the results should be presented. Although the here applied evolutionary method was seriously tested and applied to various objectives from TDGs \citep{PHR14,PRH15,Baum19}, to ram-pressure stripped DGs \citep{Steyr20}, all applying the FLASH code, and also to dEs \citep{HTG04}, we wish to roughly discuss two parameter choices for our star formation recipe: 
The influence of the cluster formation time $\tau_{cl}$ in eq. \ref{eq:SBF_thres} and the volume of a computational cell $V_{cell}$ in eq. \ref{eq:particlaMass}.
\\
Since stars are formed per numerical timestep (of the order of $10^4$ yr), the star cluster mass accumulates these temporal stellar mass fractions until a stellar particle is born after $\tau_{cl}$ representing a stellar cluster. Because these stellar mass fractions are continuously subtracted from the gas mass in each timestep, the gas mass $M_g$ decreases and by this reasonably also the SFR until $\tau_{cl}$. Due to stellar feedback (subsec. \ref{sect:feedback}) the temperature T in eq. \ref{eq:SBF_eff} even increases and enhances a decreasing SFR. 
If the $\text{SFR} = dM_s/dt = - dM_g/dt$ is proportiponal to the gas mass, $M_g(t)$ declines by $e^{-t}$, i.e. that most of a star-cluster mass is formed rapidly and after $\tau_{cl}$ $M_g$ is reduced by $e^{-\tau_{cl}}$ so that a variance by a factor of e.g. two to both sides does not change the cluster mass significantly. In fact, the value of $\tau_{cl}$ is convincingly derived as explained above.
Due to the square-dependence of the SFR in eq. \ref{eq:SBF} on the gas density, $M_g$ declines even steeper as $1/t$ (for constant T).
\\
With respect to the here presented comparison of both IMF modes it must be  emphasized that absolute values of one IMF mode play a secondary role only as long as the parametrization is the same. 
\\
Furthermore, due to dynamical effects $M_g$ in a grid cell will change during $\tau_{cl}$ but not be replenished of the same amount by gas inflow as it is reduced by star formation. Even more, the stellar feedback increases the gas pressure, acting against inflow. After $\tau_{cl}$ a new generation of cluster formation can start. 
\\
Another important measure is the grid size, which lowers or rises the cell volume $V_{cell}$ by a factor of eight for changes of the simulation resolution by a factor of two. 
As an extreme, one can ask how much a division of the whole galaxy into a few (1-4) grid cells would effect. Would it lead to a reasonable result, when all the star-forming regions (in the case of larger grid cells) are connected to a single super cluster? From observations star-forming molecular clouds are extended to scales from $<10$ to $\sim 30$ pc \citep[e.g. in NGC 300]{Faesi19} and young star clusters are separated by $30+$ pc \citep[e.g. in M51]{Grasha19}. This means that a grid resolution of $50$ pc sounds reasonable for star-cluster formation. While smaller grid cells could eventually lead to complications if HII regions get larger, doubling the scale would in reality hardly comprise a single star-forming molecular cloud so that $>1$ star clusters should form which would be artificially combined to one in the models.

\subsection{The Initial Mass Function}\label{subsec:imf}

Although the IMF implementation has already been basically described in \citet{Steyr20}, we repeat the details here in order to explain our model strategy. Once a SSP is formed, it will immediately produce feedback in the form of stellar radiation and SNeII, depending on the mass of the SSP. The IMF is a function that describes the probability for a star of mass $m$ to be born in the interval $[m,\, m+dm]$ and, by this, describes the number of newly-formed stars in each mass bin. It can be expressed by a power-law
	\begin{equation}\label{eq:imf1}
		\xi(m) = k\,m^{-\alpha}
	\end{equation}
where $k$ is a normalisation constant and $\alpha = 2.35$ \citep{Salpeter55} for a single power-law or for a multi-section power-law \citep{Kroupa01}:
     \begin{equation}
		\alpha = \left\{	\begin{array}{l l l}
			0.3 & \dots & 0.01\le m/\msun < 0.08\\
			1.3 & \dots & 0.08\le m/\msun < 0.5\\
		2.3 & \dots & 0.5\hspace{2mm} \le  m/\msun < 100
		\end{array} \right. 
     \end{equation}
The total number of stars $N_{tot}$ and the mass of the cluster $M_{cl}$ can be calculated by
\begin{align} 	
N_{tot} &= k\,\int\limits_{m_{min}}^{m_{max}} \xi(m)\, dm\label{eq:imf3.1}\\
M_{cl} &= \widetilde{k} \,\int\limits_{m_{min}}^{m_{max}} m\,\xi(m)\, dm\, . \label{eq:imf3.2} 
\end{align}

Since the IMF is a probability function, the total probability of $N_{tot}$ is unity and determines k whereas $\widetilde{k}$ is determined by the total stellar mass. For all simulations in this work, we divide the IMF into 64 equal logarithmic mass bins and apply the standard multi-section IMF by \citet{Kroupa01} for ${\xi(m)}$ in eq. \ref{eq:imf3.1} and \ref{eq:imf3.2} above.

The total number of stars $N_{tot}$ and mass $M_{cl}$ of a cluster are then given by
\begin{align}
	N_{tot} &= k_1\,\int\limits_{m_{min}}^{0.5\,\msun} m^{-1.3}\,dm + k_2\,\int\limits_{0.5\,\msun}^{m_{max}} m^{-2.3}\,dm 
\label{eq:imf_a1} \\
	M_{cl} &= k_3\,\int\limits_{m_{min}}^{0.5\,\msun} m^{-0.3}\,dm + k_4\,\int\limits_{0.5\,\msun}^{m_{max}} m^{-1.3}\,dm \, ,
\label{eq:imf_a2}
\end{align}
where $k_1, k_2$ and $k_3, k_4$ are normalisation constants evaluated via the cluster mass and must fulfil the condition
\begin{align}
k_1\,\frac{\big( 0.5\,\msun \big)^{-0.3}}{0.3} = k_2\,\frac{\big( 0.5\,\msun \big)^{-1.3}}{1.3} 
\end{align}
and
\begin{align}
k_3\,\frac{\big( 0.5\,\msun \big)^{0.7}}{0.7} = k_4\,\frac{\big( 0.5\,\msun \big)^{-0.3}}{0.3} .
\end{align}
Therefore, $k_1 = \frac{6}{13}\,k_2$ and $k_3 = \frac{14}{3}\,k_4$. 

In order to fill all mass bins according to the IMF from a minimum up to a maximum mass - in these simulations from $m_{min} = 0.1\,\msun$ to $m_{max} = 120\,\msun$ - a minimum star-cluster mass of about $M_{cl} = 10^4\,\msun$ is necessary (see sect. \ref{sec:discussion} and \citet{PHR14}). If one assumes a star-formation timescale of 1 Myr, cluster masses of less than $10^4\,\msun$ suffice for a SFR of $\le 10^{-2}\,\SFR$ at which the SFRs derived from $Ha$ and UV start to diverge \citep{Lee09}. This leads numerically to fractions of massive stars that cannot happen in nature. 

Since one can speculate that an under-populated but full-range IMF leads to energetic and chemical consequences, this paper aims at exploring how low SFRs affect the IMF and, by this, the stellar feedback by energy and chemical abundances. 
Plausibly, there are two possibilities to form stars at low SFRs. Firstly, to build up the star cluster from the low-mass proto-stellar cores to more massive stars, but to stop when the last complete star is formed and the remaining gas reservoir is incapable of allowing the formation of a single more massive star. Secondly, SF is assumed to happen stochastically according to the IMF probability \citep{Krumholz14}. 
Modelling allows an additional third and, for simplicity, most used approach, namely, to fill the IMF irrespectively of the need for integer numbers of stars in each bin, until the uppermost maximum mass limit $m_{max}$. For SFR of $\ge 10^{-2}\,\SFR$, the incompleteness of the IMF should not occur on average, but is crucial below this value.

In this paper we intend to compare an incompletely filled IMF with that of a truncation. Incompleteness means that the massive bins can be populated with fractions of stars only while the truncation of the IMF can be realized by 
\begin{align}
	N_{m_{max}} = 1 &= k_2\,\int\limits_{m_{max}}^{m_{max}+\Delta m} m^{-2.3}\, dm \, ,\label{eq:upper_mass}
\end{align}
where $\Delta m$ means the uppermost mass bin filled with the very last single star. 
It should be emphasized, that in both of our realizations, the IMF is not stochastically sampled as e.g. in \cite{App20, Smith21, JKF21, Gatto17} among others.
The sensitivity of the choice of $\Delta m$ will be discussed in subsec. \ref{sect:analyt}. 

The differences between the filled and truncated IMF, can be analyticly and numerically quantified, respectively. For the truncated IMF, the uppermost stellar mass $m_{max}$ is a function of the cluster mass $M_{cl}$ and can be evaluated by
\begin{align}
M_{cl} &= k_3\,\int\limits_{0.1\,\msun}^{0.5\,\msun} m^{-0.3}\,dm + k_4\,\int\limits_{0.5\,\msun}^{m_{max}} m^{-1.3}\,dm \label{eq:cluster_mass}\, .
\end{align}

These eq.s \ref{eq:upper_mass} and \ref{eq:cluster_mass} are solved analytically, and to now obtain $m_{max}$ for a certain $M_{cl}$, a root finding method (e.g. the bisection method) is used in the numerical code.

\subsection{Stellar Feedback}\label{sect:feedback}

In our simulations we consider feedback from different sources, as there are stellar radiation and winds, SNe type Ia and II, and AGB stars. Because the latter terminate the stellar lives much later than massive stars, the mass-dependent lifetimes of stars are also a relevant ingredient.
The lifetime of stars as a function of mass derived by \citep{RCK09} is applied (see \citet{Steyr20}). Although the treatment of stellar feedback is already documented in \citet{PHR14} and \citet{Steyr20}, we stress its formulation here again because the understanding of the different outcomes for filled and truncated IMFs can be more obviously understood by the inspection of the formulae.

\subsubsection{Stellar Radiation}\label{radiation}

To complete the self-regulation of SF by stellar feedback, stellar radiation has to be taken into account. Massive stars with a mass above $8\,\msun$ emit Lyman continuum photons with a flux $S_\ast$ that increases with mass and which completely ionizes the ISM within the Str\"omgren sphere of radius $R_S$, resulting in an HII region. In this hot environment, SF cannot happen, so it is self-regulated. The Str\"omgren sphere can be derived as the equilibrium of ionisation by Lyman continuum radiation and the recombination rate of surrounding hydrogen
	\begin{equation}\label{eq:Rs}
		R_S = \left( \frac{3\,S_\ast}{4\pi\,n^2_H\,\beta_2} \right)^{1/3}\, ,
	\end{equation}
where $S_\ast$ is the ionising photon flux, $n_H$ the hydrogen number density, and $\beta_2$ its recombination coefficient. The ionising photon flux by Lyman continuum radiation can be approximated by \citep{PRH15}
	\begin{equation}\label{eq:L_ly}
          S_\ast(m) = 3.6\times10^{42} \left(\frac{\bar m}{\msun}\right)^4\, ,
	\end{equation}
where $\bar{m}$ is the average mass within an IMF mass bin.
Depending on the resolution, the Str\"omgren radius can be smaller than a typical grid cell. Therefore a better sub-grid description is needed, where the mass of the HII region is calculated. 
The wind and radiative driven feedback (energy transfer efficiency) is determined by numerical simulations to be of the order of less than $1 \%$ only \citep{FHY03,FHY06,Hen07}. The models in these papers (e.g. for $60\,\msun$) show that the bubble expands to a radius of $12.5\,\text{pc}$ already after $10^4\,\text{yr}$ but with declining speed and thereafter suffers shell disruption by instabilities.
The temperature within the Str\"omgren sphere is set to $2\times 10^4\,\text{K}$ so that the average temperature within a grid cell is then the mass-weighted average between the Str\"omgren sphere temperature and the actual temperature of the cell.

Importantly, for the comparison of filled vs. truncated IMF the total Lyman continuum flux reads
    \begin{equation}\label{eq:Ly_tot}
 Ly_c \propto \int\limits_{8}^{m_{max}}\,S_\ast(m) \xi(m)\, dm\ ,
    \end{equation} 
so that the total radiative Lyman continuum energy of a single stellar population has to be multiplied by the stellar lifetime $\tau_*$ and amounts to 
    \begin{equation}\label{eq:E_Lyc}
       E_{Lyc} \propto \int\limits_{8}^{m_{max}}\,S_\ast \tau_*(m) \xi(m)\, dm\ . 
    \end{equation}

\subsubsection{Stellar Winds}

Massive stars also heat the star-forming sites by their winds. 
The metal-dependent mass-loss rate by winds of OB stars on the main sequence is approximated \citep{hen87,TBH92} by
	\begin{equation}\label{eq:mdot}
		\dot{m}_w = 10^{-15}\,\left(\frac{Z}{\Zsol}\right)^{1/2}\,\left(\frac{L}{L_\odot}\right)^{1.6}\,\text{M}_\odot\,\text{yr}^{-1} \, ,
	\end{equation}
where $Z$ is the stellar metallicity and the luminosity $L$ is calculated from the mass-luminosity relation \cite[table 1]{Maeder96}. Within a numerical timestep $\Delta t$, therefore, a stellar mass 
     \begin{equation}
        m_{w,\Delta} = \dot{m}_w \cdot \Delta t 
     \end{equation}
is released by winds as long as the time since the formation of the SSP is shorter than the stellar lifetime $\tau_\ast(m)$. 
To obtain the total wind mass loss of a specific mass bin, for which a mean lifetime $\bar{\tau}_\ast$ of all stars in that bin and the assumption of a constant wind rate $\dot{m}_w$ over the lifetime can be applied (but see e.g.  \citet{FHY03}), one can calculate 
     \begin{equation}
    M_{w,bin} = \dot{m}_w \cdot \bar{\tau}_\ast \cdot N_{\ast,bin} \, .
     \end{equation}
Hence, all the $N_{\ast,bin}$ stars in a specific mass bin with the average bin mass $\bar{m}$ lose a wind mass fraction of the bin
   \begin{equation}
      f_{m,w} = \frac{M_{w,bin}}{\bar{m} \cdot N_{\ast,bin}} \, .
   \end{equation}

For the wind power $\dot{E}$ exerted to the ISM, the final wind velocity for a star within an IMF mass bin of average mass $\bar{m}$ is then given by
	\begin{equation}\label{eq:v_inf}
		v_\infty = 3\times10^3\,\left(\frac{\bar{m}}{M_\odot}\right)^{0.15}\,\left(\frac{Z}{\Zsol}\right)^{0.08}\,\text{km}\, \text{s}^{-1}\, .
	\end{equation}
As for eq. \ref{eq:mdot}, this dependence is derived by \cite{hen87} from the massive-star atmosphere and wind models by \cite{KPP87}.
The heating of the ISM by the OB stellar winds results from their kinetic power and can be expressed for each single star as
	\begin{equation}\label{eq:E_kin}
		\dot{E}_{kin} = \frac{1}{2}\,\dot{m}_w \,v_\infty^2 .
	\end{equation}
Practically, the total wind energy of $N_\ast$ stars of all mass bins populated with stars up to $m_{max}$ during a numerical timestep $\Delta t$ is accumulated by 
	\begin{equation}\label{eq:Ewinddt}
	E_{w,\Delta t} = \int\limits_8^{m_{max}} \frac{dE_w(m)}{dt} N_\ast (m)\, dm \cdot \Delta t
	\end{equation}
While stellar radiative heating is included for the formation of HII regions around massive stars - not spatially resolved, but analytically implied as subgrid physics within the cells - the energy release by radiation + wind to the surrounding ISM within a cell can be calculated according to eq. \ref{eq:E_Lyc} and \ref{eq:Ewinddt}, but reduced by an energy transfer efficiency $\epsilon_w$ of $5\%$. This value is arbitrarily assumed although \cite{Hen07} derived smaller values from their models \citep[Danica Kroeger, private comm. 2006]{FHY03,FHY06}.

\subsubsection{Type II Supernovae}\label{subsec:SNIIfb}

Stars with masses larger than $m_\ast = 8\,\msun$ end their lives as core-collapse SNeII. The feedback taken into account is purely thermal, but the SN remnants expand due to their overpressure. 
For each SNII explosion an ejected energy of $10^{51}$ erg is applied, but reduced by an efficiency of $5\,\%$, i.e. $5\times10^{49}$ erg per SNII is transferred into the ISM. 
This value is arbitrarily chosen and in the realistic range of $10\,\%$ from single 1D explosions models \citep{TGJ98}, $2\,\%$ derived from superbubble-to-galactic winds in simulations \citep{RH13} and $1-3\,\%$ from recent models of supernova-driven ISM turbulence  \citep{CS20}. 
When both gas phases are allowed to mix, the cooling timescale for the mean temperature at the total gas density is too short to provide any energetic effect on the environmental ISM. To avoid this well-known overcooling problem, we use the recipe applied by \citet{PHR14}. 

The SNII rate for a SSP can be easily expressed as an analytical function of the total stellar cluster mass according to the IMF and the stellar lifetimes. In computational applications of a SSP, the number of SNeII from a mass bin is given by the number of massive stars in that bin whose lifetime $\bar{\tau}_\ast$ is reached by the cluster age.
The mass of SNeII in a mass bin with mean mass $\bar{m}$ then depends on the mass fraction remaining from the wind loss, and is reduced by the remnant mass, which ranges between $1.3$ and $2.1\,\msun$ for initially massive progenitors: 
\begin{align}
f_{m,rem} &= \frac{m_{rem}}{\bar{m}}\, , \\
f_{m,SNeII} &= 1 - f_{m,rem} - f_{m,w} \, .
\end{align}
From this the total SNII mass loss of a bin results from $f_{m,SNeII}$ multiplied by $N_{\ast,bin}$ and $\bar{m}$.

\subsubsection{Asymptotic Giant Branch Stars} 

Stars with masses below $m_\ast = 8\,\msun$ return mass to the ISM during the asymptotic giant branch (AGB) phase and end as white dwarfs (WDs), assuming a mean WD mass of $0.6\,\msun$. A fraction of these WDs will terminate their life as SNIa according to the next subsection. The ejected mass comprises processed elements (see section \ref{subsec:yields}), while the energetics of AGBs winds is neglected.

\subsubsection{Type Ia Supernovae}

SNeIa are the final explosions of a C-O WD when it exceeds the Chandrasekhar mass ($1.44\,\msun$) by means of mass accretion from its companion in a binary system or through their merging. The SNIa rate depends on the mass ratio $\mu_i$ of the primary or the secondary star with respect to the binary mass. Its distribution function and the probability of a star to be secondary in a binary system is described in \citet{Steyr20}.

The number of SNIa $N_{SNIa}$ in each mass bin can be calculated. The SNIa returns the complete final binary mass to the ISM with an energy efficiency of $5\%$ of $10^{51}$ erg as for SNeII 
and a delay time after the star formation of about 60 Myr \cite[see fig. 9]{PHR14} in agreement with \cite{Greggio10} and slightly shorter than values found by \cite{MMB12}. 
Since SNeIa act as single explosion, we suggest that they stir-up the ISM more effectively than SNII explosions.

\subsubsection{Chemical Feedback}\label{subsec:yields}

At the end of its lifetime, a star not only returns mass in general to the ISM, but the material also comprises nucleosynthesis products, which lead to an enrichment of heavier elements as chemical feedback. SNeII enrich their surroundings mostly with $\alpha$-elements like Ne and O, whereas SNeIa yield mostly Fe taken from the W7 model of \citet{Trav04}. For intermediate-mass stars their enrichment during the AGB phase is dominated by N and C. 
We use the stellar yields from \citet{MBC96} for stellar masses $m_\ast = (1 - 4)\,\msun$ and \citet{PCB98} for masses above $6\,\msun$ with a linear interpolation between $4$ and $6\,\msun$.

\section{Results}\label{sec:results}

\subsection{The Evolution}

\begin{figure*}
\includegraphics[width=\columnwidth]{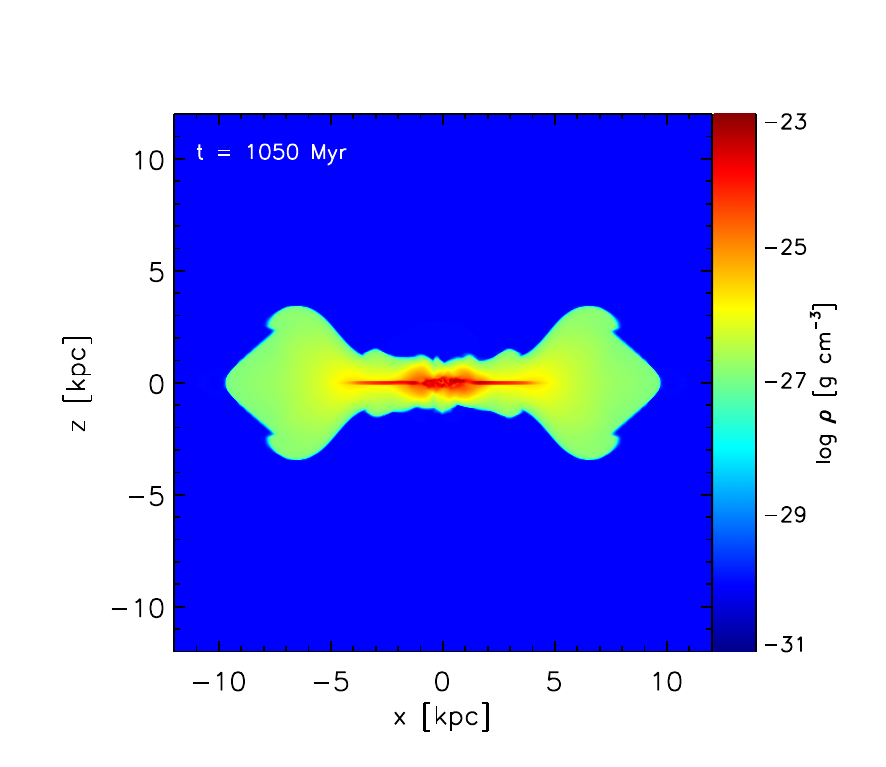} 
\includegraphics[width=\columnwidth]{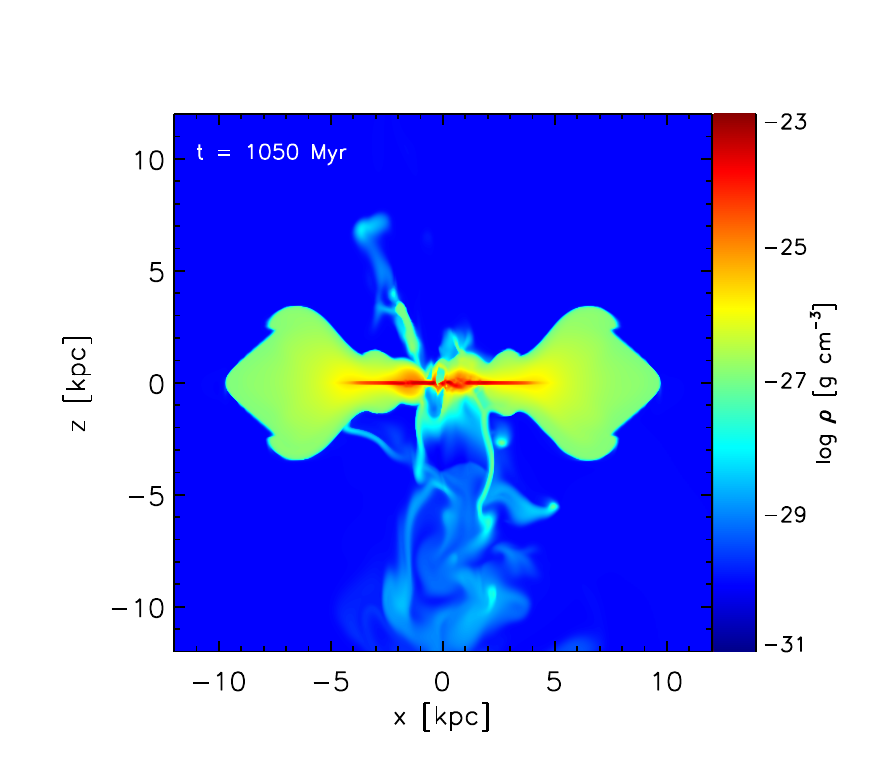}
	\caption{Edge-on slice through the filled IMF model (left) and truncated IMF model (right) at the same simulation time of $t _{sim}= 1050\,\text{Myr}$. The truncated IMF exhibits a strong bipolar outflow due to a higher SFR and a locally weaker feedback compared to the filled IMF. The colour bar is given in logarithmic volume densities.}
	\label{fig:denscut}
\end{figure*}

Fig. \ref{fig:denscut} shows a snapshot at a simulation time of $t_{sim}=1050\,\text{Myr}$ for both the filled and truncated IMF, respectively, as a density cut through the $xz$-plane at $y = 0$.
The highest spatial resolution is in both cases $50\,\text{pc}$ in each dimension. The truncated IMF (right panel) exhibits a strong galactic bipolar outflow, driven by the larger SFR and SNeII (see next sect. and sect. \ref{sec:SNII}) rate. With velocities of up to $600\,\text{km}\,\text{s}^{-1}$ a galactic wind is driven. The filled IMF on the left-hand side does not show such an outflow.

\subsection{Star-Formation Rate}

\begin{figure}
\includegraphics[width=\columnwidth]{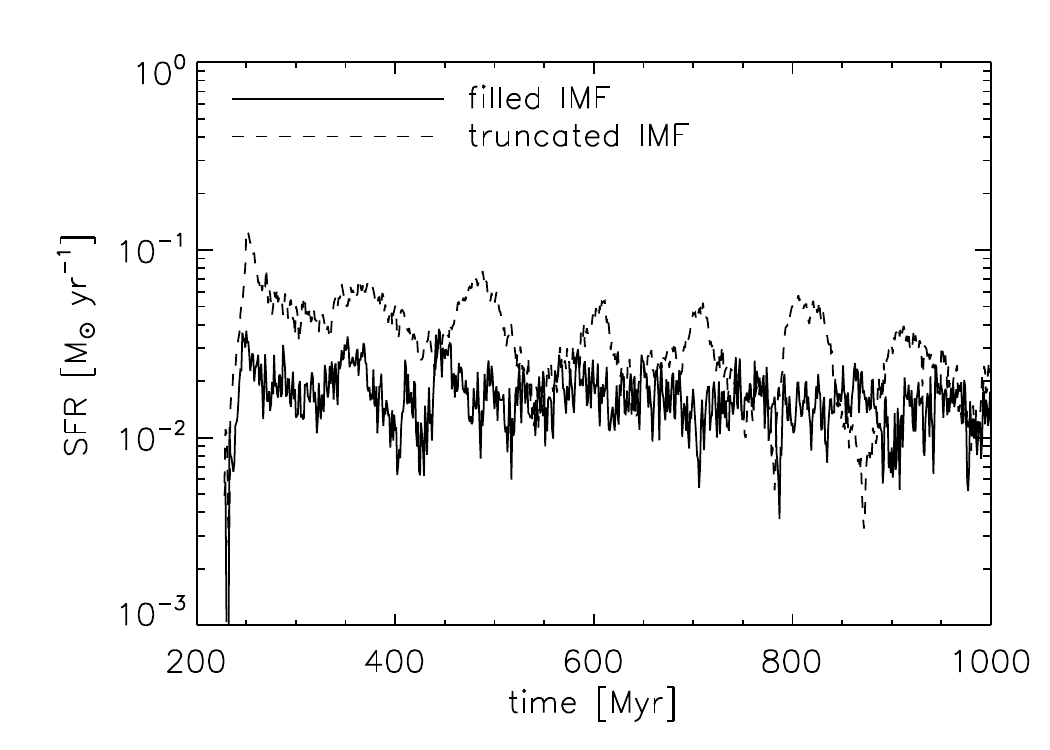}
	\caption{SFR in 1 Myr bins for the filled (solid line) and truncated (dashed line) IMF, respectively.}
	\label{fig:sfr}
\end{figure}

The SFR, is the most important process for the evolution of galaxies and is depicted in fig. \ref{fig:sfr} for both IMF modes, filled (solid line) and truncated (dashed line).

If the total SFR of a model exceeds the critical value of $10^{-2} \, \SFR$ the formation of individual star clusters can fall beneath the critical mass for a full IMF (\citet{PHR14}). 
In the case of the filled IMF, the actual presence of even numerical fractions in massive star bins allows the regulation of SF by feedback of all massive stars instantaneously. 
Importantly, the radiative energy release (Lyman continuum) depends on the stellar mass by a steeper positive power than the absolute value of the negative IMF slope (see eq. \ref{eq:L_ly}). 
This leads additionally to a somehow over-regulated SF with respect to the truncated IMF, resulting in a lower SFR as already seen in \citet{PHR14}. 

Due to the lack of the most massive stars, the radiative stellar feedback of the truncated IMF is smaller and SNII explosions are delayed due to the longer lifetime of the $m_{max}$ stars and the lack of very short-lived stars. Both result in an initially larger SFR. The SF of the truncated IMF is further characterized by oscillations (see fig. \ref{fig:sfr}) with periods around $120 \,\text{Myr}$, almost the free-fall time of the DG. This so-called "breathing" happens here also for a flattened rotating disky DG, while \cite{Schroy11} discover its occurrence in non-rotating DG models.

\subsection{Supernova type II Feedback}\label{sec:SNII}

\begin{figure}
	\includegraphics[width=\columnwidth]{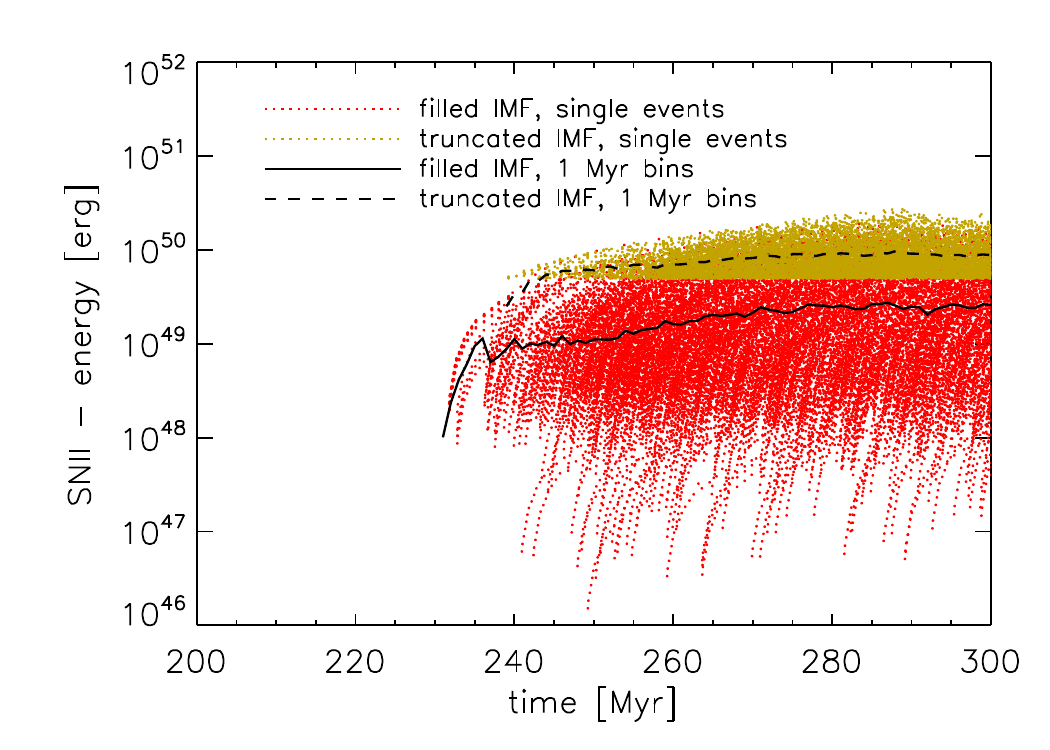}
	\caption{SNII energy (5\% of the total SNII energy of $10^{51}\,\text{erg}$) for the filled IMF (red dots and solid black line) and the truncated IMF (yellow dots and dashed black line). The lines are the SNII energy in $1\,\text{Myr}$ bins and the dots represent the SNII energy for a single cluster at the current time step. Shown are the first $80\,\text{Myr}$ of feedback from type II SNe.}
	\label{fig:SNenergy}
\end{figure}

Fig. \ref{fig:SNenergy} shows the SNII energies for both filled and truncated IMF, respectively, at the first 70 Myr of star formation. 
The solid and dashed lines indicate the average SNII energy per Myr for the filled and truncated IMF, respectively.
The dots (red for the filled and yellow for the truncated IMF) represent the SNII energy of a single stellar cluster in the current timesteps. Chained dots towards the upper-right direction, most clearly discernible for the filled IMF (red), belong to the same star cluster and reveal the increasing SNII number with the aging of the SSP. That many cluster SNeII start mostly below $10^{47}\,\text{erg}$ demonstrates that from massive stars only small fractions of SNII energy is released.  
The lower cut-off at $5 \times 10^{49}\,\text{erg}$ for the truncated IMF reflects directly the SNII energy transfer efficiency of $5\%$. 

This leads to lower temperatures of the hot expelled gas of about $10^5\,\text{K}$ that cools more efficiently, which means that the thermal feedback by SNeII is reduced, even with the cooling procedure mentioned above.
The thermal energy is thus too low and is efficiently cooled away to drive a galactic wind.

In contrast, the truncated model allows only explosions for integer SNII numbers in each mass bin, i.e. with more energetic SN power for each individual explosion. This can explain why a massive galactic wind is driven (right-hand fig. \ref{fig:denscut}). At this stage it cannot be definitively concluded from the models which of both factors contributes more to the galactic wind, the stronger SFR or the larger SNII  energy, or both. We will focus on this question analytically in sect. \ref{sect:analyt}. 

As one can also discern from fig. \ref{fig:SNenergy}, the SNII feedback starts earlier for the filled IMF, because though only fractions, the lifetimes of the most massive stars are shortest. They explode already after a few Myr, while for the truncated IMF the SNeII are delayed.

\subsection{Cluster Mass Function}

\begin{figure}
	\includegraphics[width=\columnwidth]{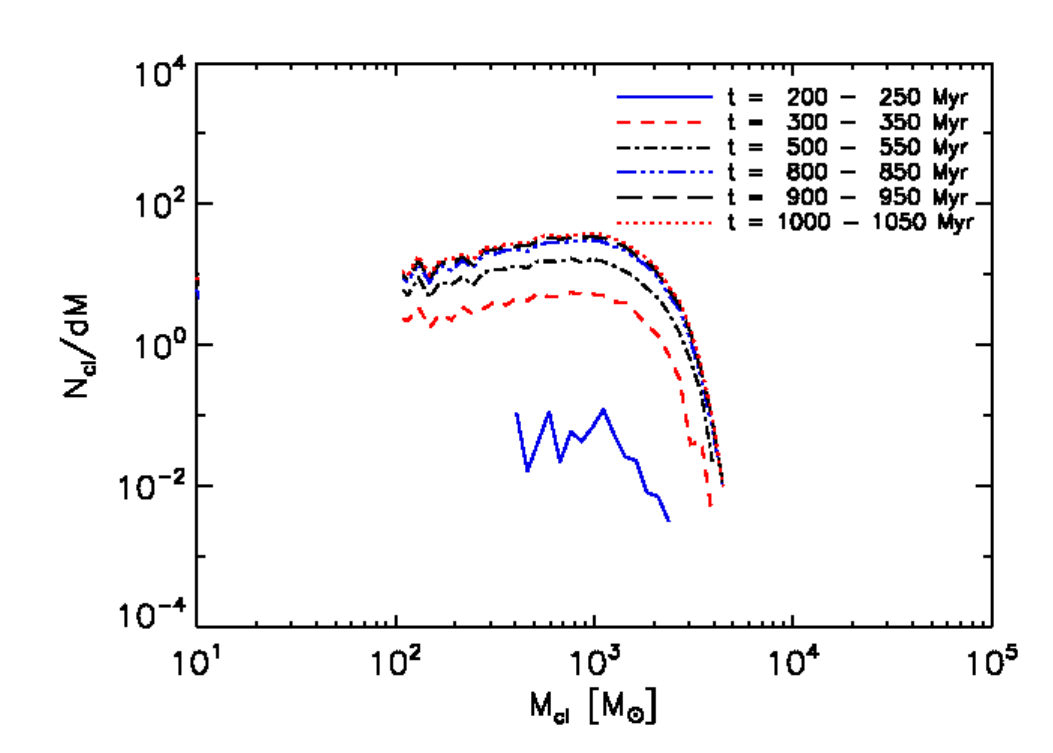}
	\includegraphics[width=\columnwidth]{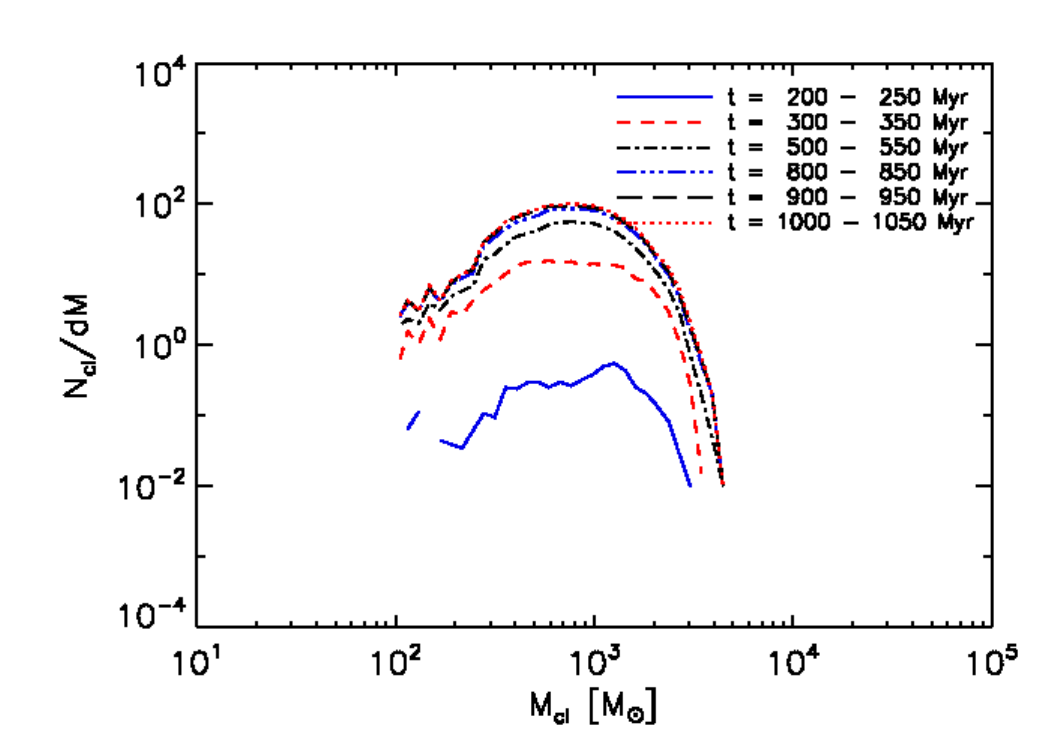}
	\caption{The cluster mass function in 50 Myr age bins for the filled IMF (top panel) and the truncated IMF (bottom panel) shows the number of clusters per initial mass bin (dM) as a function of the cluster mass.}
	\label{fig:cmf}
\end{figure}

Although the global SFR stays mostly above $10^{-2} \SFR$, i.e. filling the IMF in the case of a single star cluster, the cluster mass function (CMF) in fig. \ref{fig:cmf} shows cluster masses always below $10^4 \msun$, which is insufficient to complete the IMF.

For the CMF for the filled and truncated IMF, respectively, one can notice, that the truncated IMF has more massive clusters than the filled IMF. The differences between the CMFs already start quite early, during the first $30\,\text{Myr}$ of SF. 
The reasons for these are manyfold: As already mentioned, the SF of the filled IMF gets more strongly regulated by the feedback of massive stars and with a larger power of released energy (see sec. \ref{radiation}) even for fractions of massive stars. Clusters with the truncated IMF have more time to accumulate mass within one Myr of cluster-formation time until they are capable of producing entire massive stars, which then start to regulate the SF by their feedback. Therefore, the truncated IMF initially forms more star clusters, resulting in a higher SFR at the beginning. 
After a simulation time of 250 Myr the difference in the CMF gets more pronounced, and the truncated IMF produces more massive clusters. Surprisingly, both peak at a cluster mass of $M_{cl} = 10^3\,\msun$. The difference is not that large for both CMFs in the amplitude of the peak, but in their shape. 

That the CMF deviates from the global shape, must be interpreted as follows: With respect to the standard CMF slope $\frac{dN}{dM_{ecl}} \propto M_{ecl}^\beta$ of $\beta = -2$, at lower cluster masses the CMF flattens or even decreases \citep{deGPL05,Larsen09} but in the massive range of $> 10^4\,\msun$, steepens to $-2.7$ and higher, e.g. \citet[for M83]{WS21}. 
The compilation of CMFs from the recent literature by \cite{Krum19}, however, does not show a clear tendency of slopes from massive to less massive cluster ranges. This result supports any conclusion from the observations by \cite{HED03} of $\beta \simeq -2.4$ for the Large and Small Magellanic Clouds (LMC and SMC) and \cite{deGA06} of $\beta =-1.8\pm 0.1$ (LMC) and $-2.00\pm 0.15$ (SMC) that the CMF could be time-dependent. 
Moreover, the most important caveat for the comparison of our CMF with those literature values is that observed CMFs hardly reach below $M_{cl} \sim 10^4\,\msun$. Because of the low SFR in our simulations, clusters could also not gain such large masses as it was the case for TDGs where gas inflow enhances the star formation \citep{PHR14}. This supports the $M_{cl,max}$-SFR correlation discussed before.

\subsection{Chemical Abundances} 

\begin{figure}
	\includegraphics[width=\columnwidth]{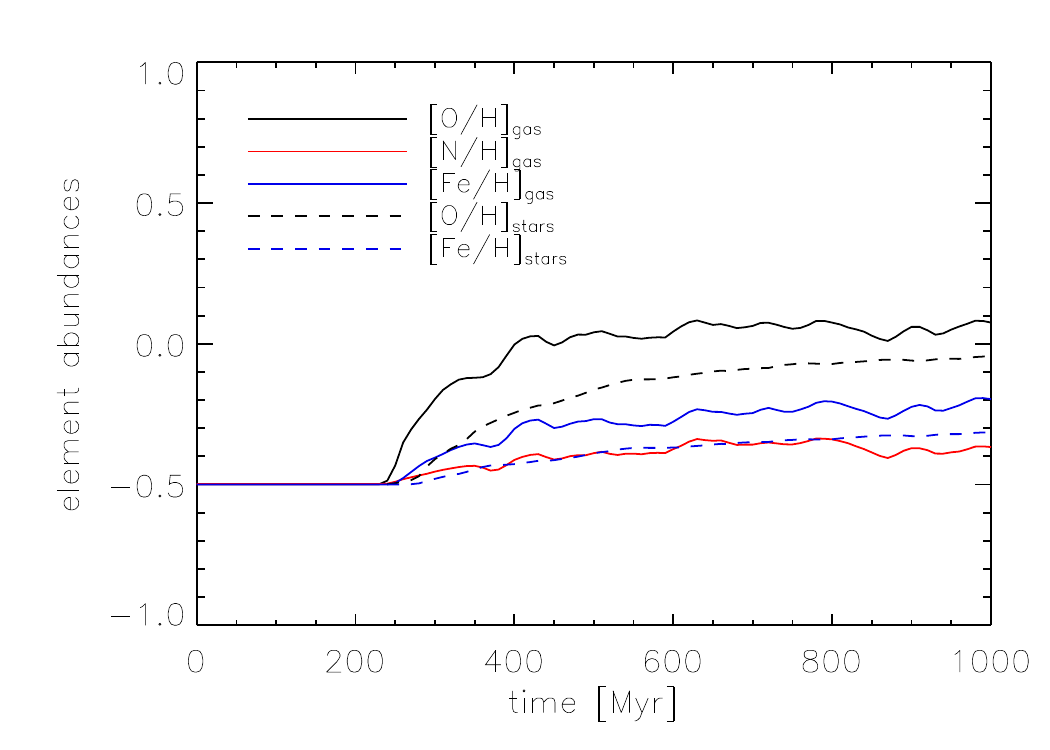}
	\includegraphics[width=\columnwidth]{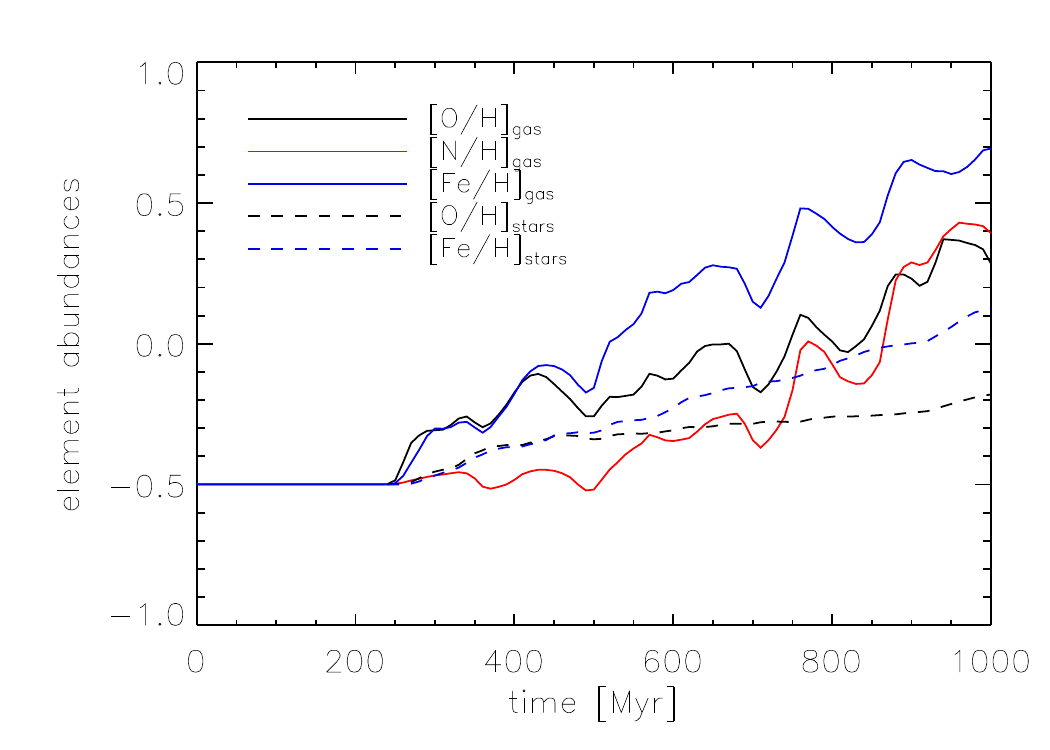}
	\caption{Element abundances for the gaseous (solid lines) and the stellar component (dashed lines) within the central region as a function of time. Top panel: filled IMF, bottom panel: truncated IMF. }
	\label{fig:ELabund2}
\end{figure}

\begin{figure*}
	\vspace{-0.5cm}
	\includegraphics[width=3in]{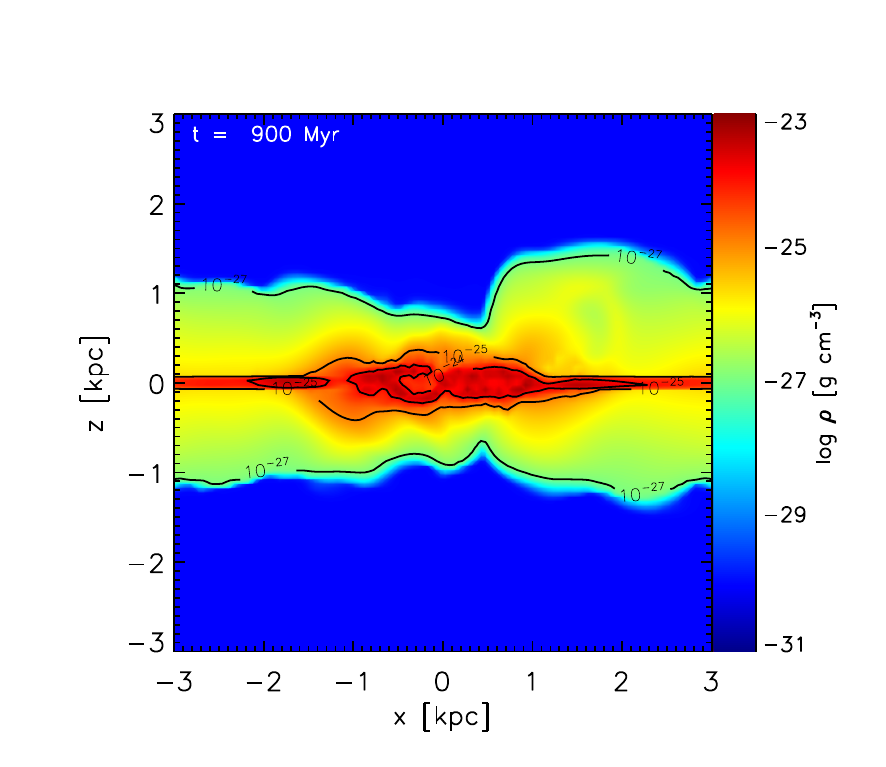} \includegraphics[width=3in]{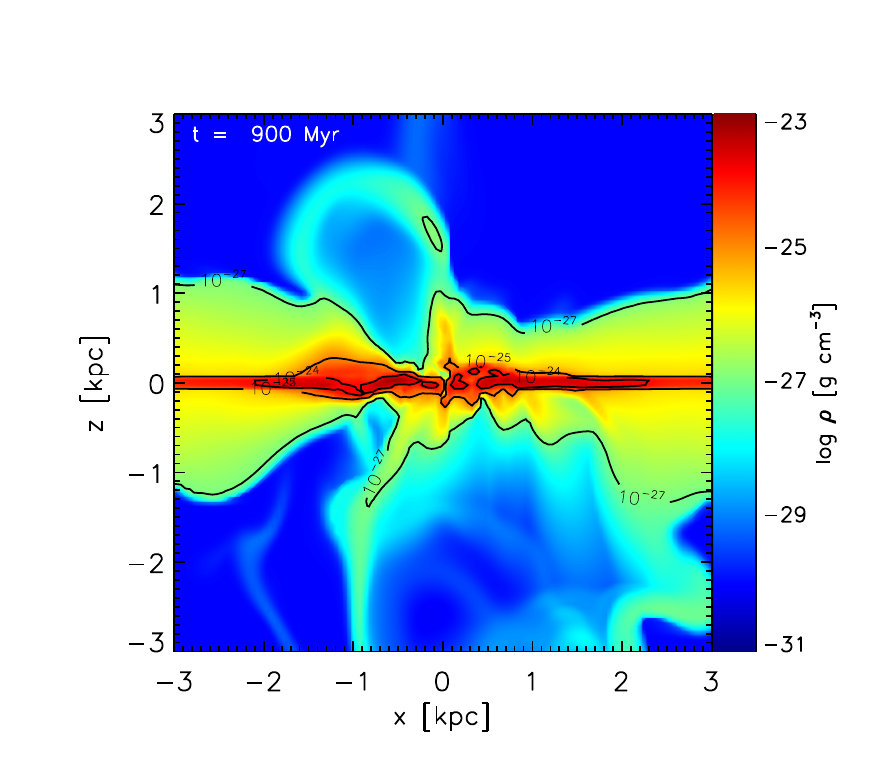} \\
	\vspace{-1cm}
	\includegraphics[width=3in]{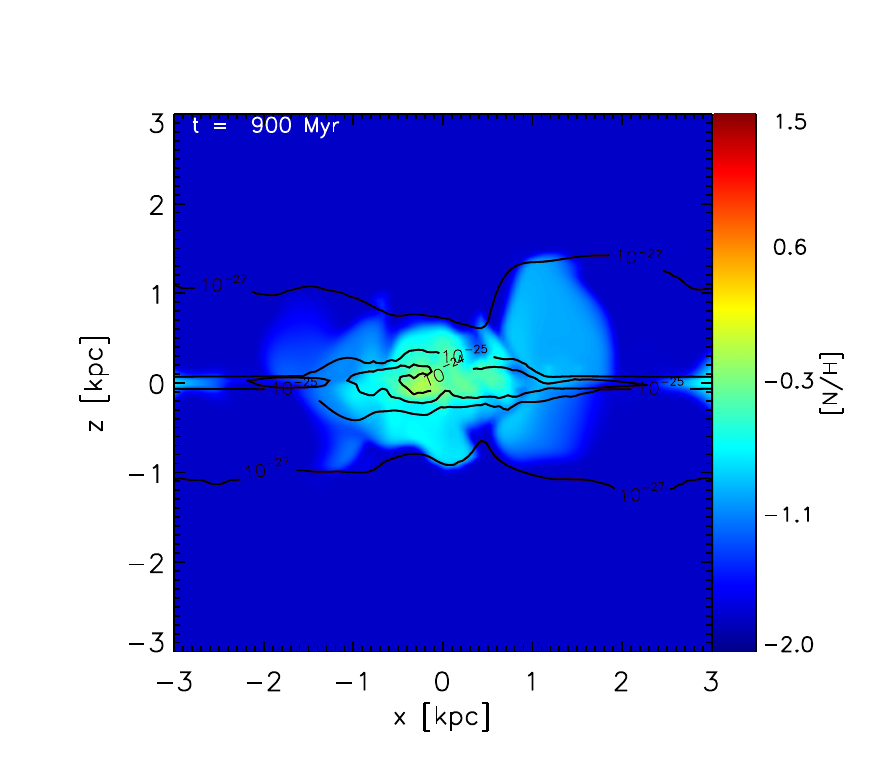} \includegraphics[width=3in]{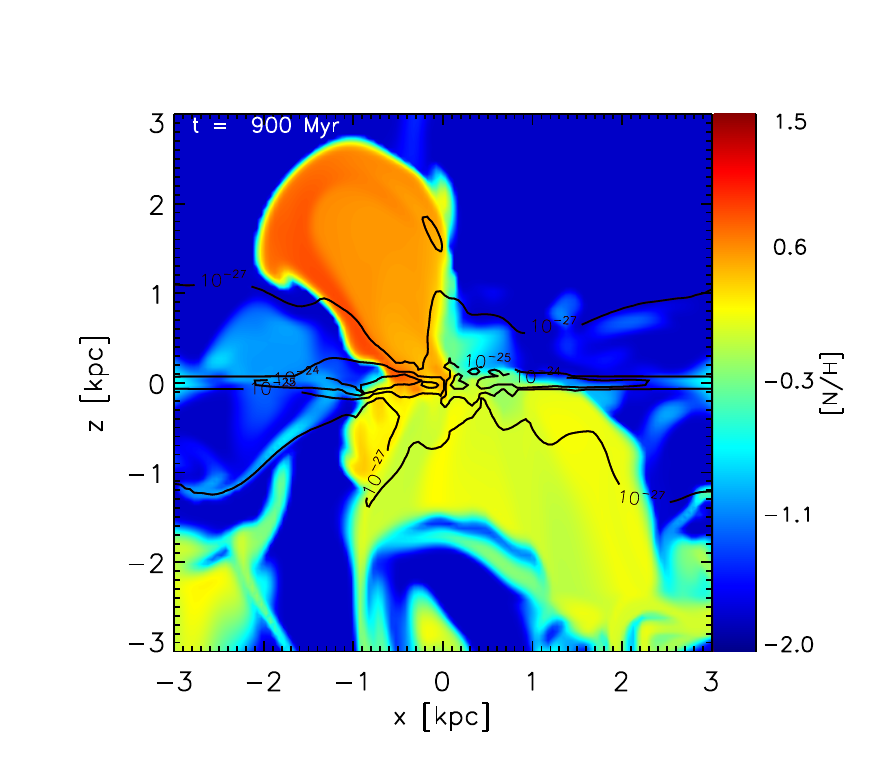} \\
	\vspace{-1cm}
	\includegraphics[width=3in]{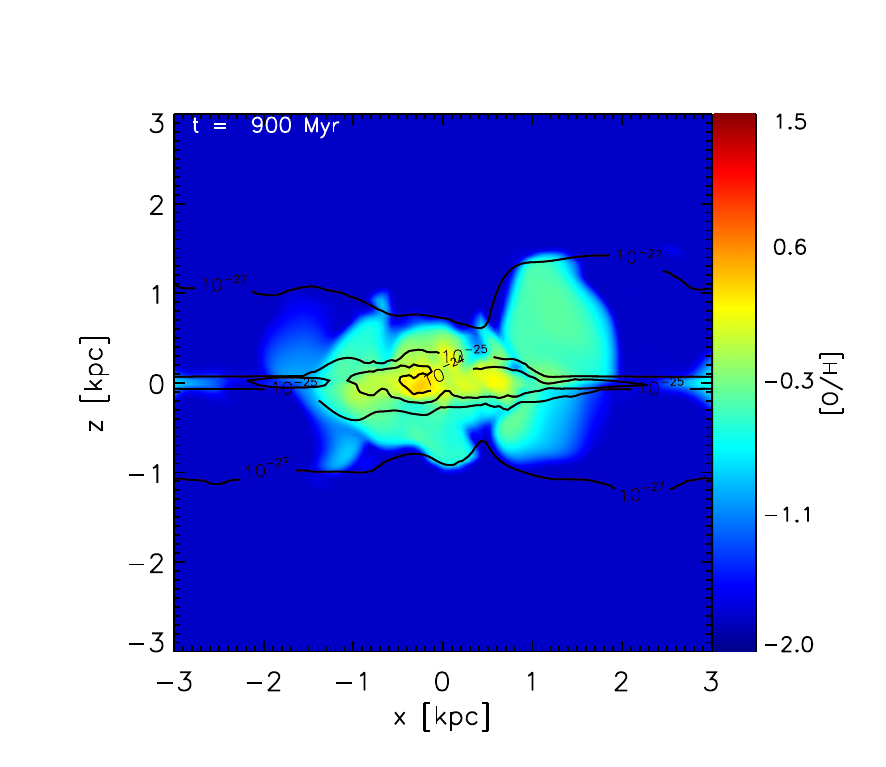} \includegraphics[width=3in]{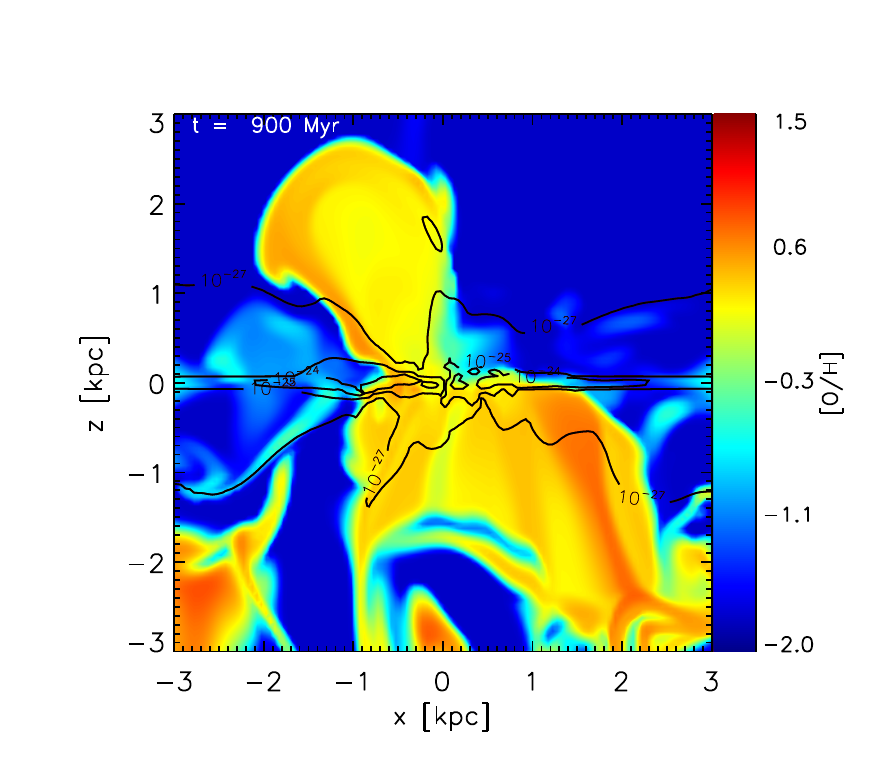} \\
	\vspace{-1cm}
	\includegraphics[width=3in]{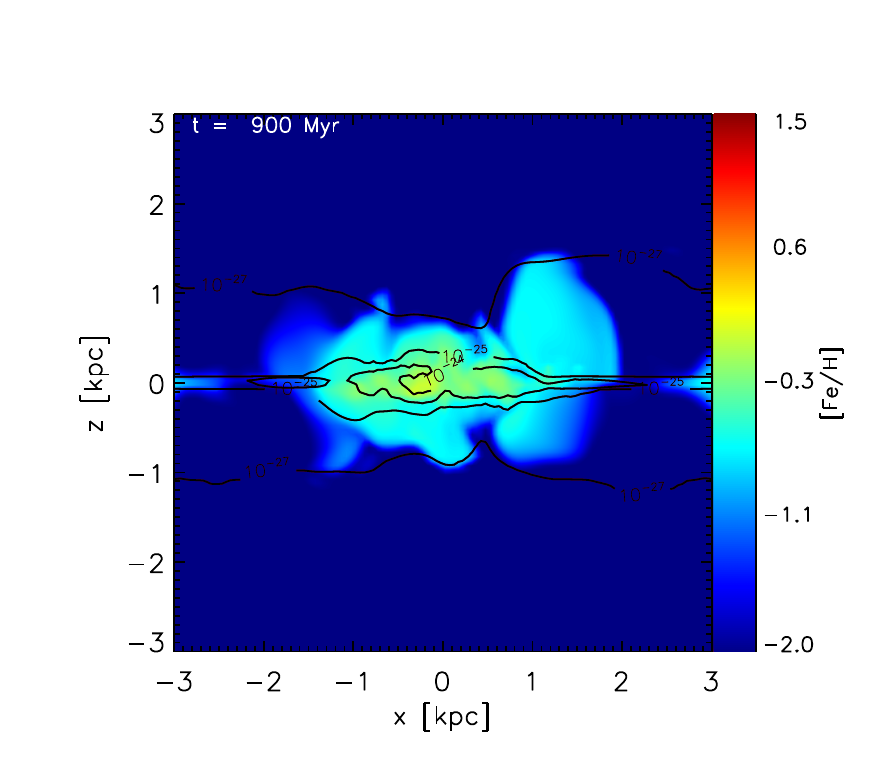} \includegraphics[width=3in]{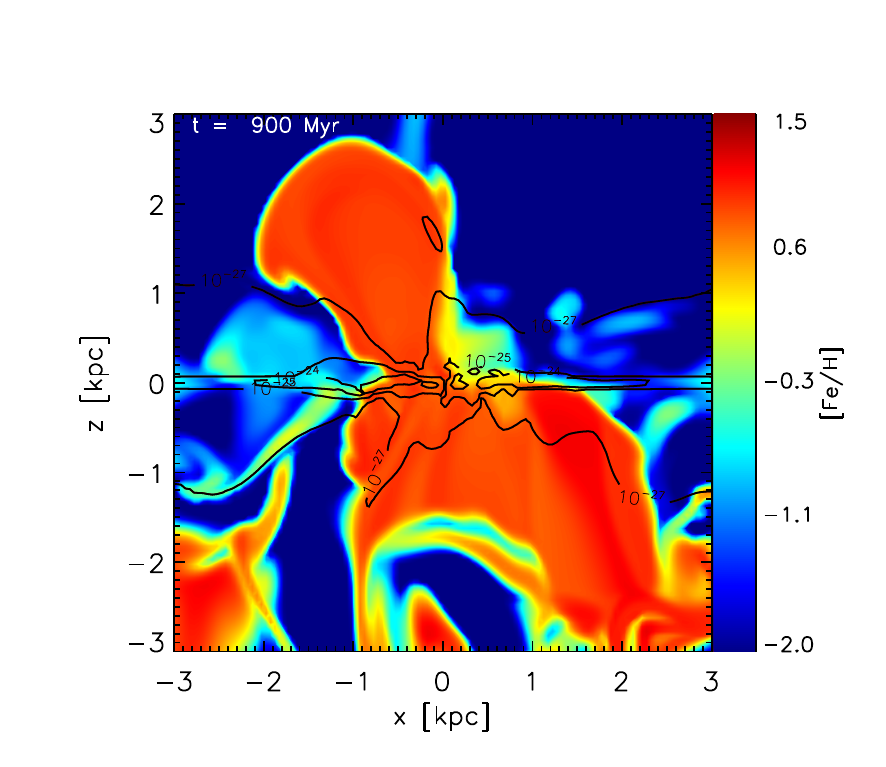}
	\caption{Slices through the $xz$-plane at $y=0$ at a simulation time of $t = 900\,\text{Myr}$ for the filled (left) and truncated (right) IMF, respectively. From top to bottom: density in $\log\,[\text{g}\,\,\text{cm}^{-3}]$, [N/H], [O/H] and [Fe/H]. Overlaid are density contours for $10^{-27},\,10^{-25},\,10^{-24}\,\text{g}\,\,\text{cm}^{-3}$}
	\label{fig:ELabund3}
\end{figure*}

In order to interpret deviations from a global IMF, abundance ratios of characteristic chemical elements are a crucial signature. 
Reasonably, the evolution of the chemical abundances between the filled and truncated IMF differ. Fig. \ref{fig:ELabund2} shows the time evolution of N, O, and Fe and their abundance ratios with an initial metallicity of $Z=0.32\,\Zsol$ as averaged values within a cylinder of $r = 0.5\,\text{kpc}$ and hight $z = 0.2\,\text{kpc}$.  
The abundance ratio between two elements $X_i$ is defined as
\begin{equation}
\left[\frac{X_1}{X_2}\right] = \log_{10}\left(\frac{X_1}{X_2}\right) - \log_{10}\left(\frac{X_1}{X_2}\right)_\odot\,.
\end{equation}
 
There are some main properties when comparing the filled and the truncated IMF: \\
(i) In the case of the filled IMF (fig. \ref{fig:ELabund2}, top panel), the element abundances approach constant values, while those for the truncated IMF (bottom panel) increase. \\
(ii) The [O/Fe] value of the ISM follows the typical SNII yields of $\sim 0.3$, because at these short timescale the SNIa enrichment is delayed and still minor. \\
(iii) The O and Fe abundance experience an unusal steep rise after the first stellar generation injects its stellar products. This can be understood by a rough estimate: Since the abundance determinations are for both, gas and stars, the volume under consideration is limited to the star-forming central region, a flat central puck of $\sim 1$ kpc radius and 100 pc height. With the density of $1\,\text{cm}^{-3}$ almost a million $\msun$ are contained in this puck. 
A SFR of $10^{-2}\,\msun\,\text{yr}^{-1}$ over $10^7$ yr, i.e. the lifetime of about a $20\,\msun$ star, allows the use of up to $10\%$ of the gas, while the most massive stars are already injecting their chemical products. The continuation of star formation must therefore be supported by gas instream with the same initial metallicity of $1/3\,\text{Z}_\odot$, i.e. an initial O content of $10^6\,\msun\,\times\, 3\cdot 10^{-3}$ O mass fraction. 
For a full IMF with $10^5\,\msun$, almost $1/10$ of the total stellar mass is processed and returned over the stellar lifetime of $2 - 3\cdot 10^7$ yr and of this, $10\%$ is O-rich, which results in $10^3\,\msun$ oxygen or an increase of log [O/H] to $-0.35$. After $10^8$ years, $1\%$ of the stellar mass returned already amounts to $10^4\,\msun$ of oxygen and leads to a further growth, but with dilution due to dynamical effects. The differences between the IMF models are directly discernible in fig. \ref{fig:ELabund3}.
\\
(iv) The truncated IMF model starts with the same O and Fe enrichment, but has a slightly earlier Fe contribution. This can be understood from the mass truncation so that the larger O production normalized to the IMF mass from the more massive stars, does not contribute here as it does for the full IMF.
\\
(v) Well understood for both models independently, N remains below Fe and follows its enhancement with the same curvature, but very intriguingly the ratio of [N/O] differs clearly: it has almost constant values of about $-0.5$, while the truncated model varies strongly between $-0.3$ to even positive ratios. 

While DGs follow the extension of a regular mass-metallicity relation down to low masses \citep{Kirby13} but with increasing scatter, particular element abundances and abundance ratios in DGs show a large variety, as e.g. sub-solar [$\alpha$/Fe] in dwarf spheroidals and ultra-diffuse galaxies.

The reason for this behaviour could be, that the truncated IMF has, by definition, on average more integer massive stars, producing more SNeII which inject, therefore, more oxygen into the ISM. The filled IMF has, as already mentioned, fractions of massive star, and therefore will also inject fractions of stellar element abundances into the IMF by SNeII. 
Fig. \ref{fig:ELabund3} shows a slice through of the $xz$-plane of the volume density, [N/H], [O/H] and [Fe/H] of the filled and truncated IMF. One can clearly see that the truncated IMF produces large superbubbles, that break out of the plane and are filled with higher element abundances. These reach values up to $1.5$ of solar (in the case for [Fe/H]).

\section{Discussion}\label{sec:discussion}

\subsection{Numerical Models}

The comparison between DG simulations with filled vs. truncated IMF,  demonstrates the importance of handling the IMF in numerical simulations reasonably with regard to the SFR and the resulting cluster masses. Even under the assumption of an empirical IMF slope, low SFRs must guarantee the completeness of the stellar mass distribution i.e. the existence of integer star numbers as, e.g., by truncating the IMF or by random sampling. 
\citet{PHR14} and \citet{Elme09} have independently demonstrated that a minimum cluster mass of about $10^4 \msun$ is required to fill the IMF according to the ''universal'' law, whereas below (e.g. $10^3 \msun$) it leads to a lack of the most massive stars and, by this, a deviation in the slope of the massive IMF range.
Its relevance is caused by mainly two reasons: 
(i) While for a filled IMF the energy release by massive-star radiation sets in immediately and to a stronger power by even fractions of the most massive stars, in the truncated-IMF case the radiative feedback stems from the lower stellar mass in the massive range. By this, the SF self-regulation is stronger in the filled mode. 
(ii) With a filled IMF one can get fractions of massive stars which lead to a lower average SNII energy (see fig.\ref{fig:SNenergy}) injected into the ISM, whereas with a truncated IMF one ensures that only integer numbers of stars exist, with a higher mean mass. Even if the IMF of a star cluster is filled, its accumulated power of SNII explosions can drop beneath that of full stars in the truncated mode because their start is delayed and their full energy is released on a shorter timescale. 

This result of our simulations has to be considered with serious care because the SFRs are affected by the different IMF recipes in the sense that the lower in-situ self-regulation in the truncated IMF case inherently experiences a larger SFR. This effect can also be illustrated by fig. \ref{fig:cmf} where a truncated IMF produces more massive star clusters than a filled IMF.

Due to the formation of more massive stars in our model, this truncated IMF therefore also produces a larger SNII feedback, as discussed above, which drives a galactic wind. A normalization to equal SFRs allows a serious comparison of the IMF effects on the energetic feedback and will be presented in the next section.

The neglect of this energetic effect on the SF self-regulation could lead to major artefacts in numerical simulations of DGs when a filled IMF at low SFRs and too small cluster masses produces an unrealistic energy feedback. Since this IMF problem is irrelevant in many numerical simulations of massive galaxies, when high SFRs also lead to massive star clusters, this effect does not exist in low-mass DG models. 
Here the per mill fractions of SNII energy have the same effect as the well-known overcooling problem, because a temperature much lower than hot SN gas at about $10^4-10^5\,\text{K}$ can efficiently be cooled off. Mostly, the lack of sufficient SNII energy, e.g. in cases where modelers wish to drive a galactic wind artificially so as to solve the cusp-core problem in DGs according to the observed cored DM distribution \citep{Gov12}, is commonly compensated by unrealisticly high SN efficiencies (sometimes up to $100\,\%$) or an inconsistently high SFR. \citet{ATM11} state that in their simulations a total SNII feedback (with no SN efficiency) of $100\,\%$, even up to $500\,\%$, leads to a more realistic bulge-to-disc ratio. 

The herewith questioned full-IMF recipe explains the divergence of $Ha$ vs. UV brightnesses at SFRs below $10^{-2}\,\SFR$, e.g. \citet{Lee09}. 
If one considers that the SFR of DGs can reach even the order of $10^{-4} - 10^{-2}\,\SFR$, which leads to cluster masses that probably will not fill the IMF, the pure scope of DGs survival and their $M/L$ ratio will also be affected. 
A truncated IMF provides both, a more realistic description of the IMF depending on the cluster mass and SFR, as well as a SN energy transfer efficiency, mostly between $5 - 15\,\%$ as derived from single SN explosion models \citep{TGJ98}, and not additionally reduced by a shortage of SN energy due to fractions. 

The application of a truncated IMF will be quite useful when simulating e.g. ram-pressure stripped (RPS) DGs, firstly because of their low masses and low SFRs and secondly, when investigating star-forming RPS gas. \citet{HSN10, FGS11, YGF13, JCC14, KGJ14, JKK15} derive active or recent SF in such gas tails with rates between $10^{-4} - 10^{-2}\,\SFR$.
Moreover, the frequently doubted survival of DGs formed in the tidal tails of interacting systems which also show very low SFRs (e.g. \citet{VMK08}) must be newly considered under this aspect \citep{PHR14}. 
And finally, the extrapolation of the past SFR history in early-type DGs and the derivation of energies and chemical abundances, requiring e.g. strong galactic winds in order to reduce the effective yield, must be reconsidered. 
As demonstrated, the chemical abundances are also affected by the implication of a truncated IMF. Instead of artificial assumptions such as e.g. differential winds \citep{Rec08} to understand low $\alpha$-element abundances in low-mass galaxies, variations of the IMF should be taken into considerations.

\subsection{An Analytic Comparison between filled vs. truncated IMF} 
\label{sect:analyt}

Based on the two-fold IMF (eqs. \ref{eq:imf_a1} and \ref{eq:imf_a2}), we wish here to obtain a general insight as to how truncated vs. filled IMF of a specified cluster mass, i.e. the upper-mass truncation, changes the energetic feedback. This was not yet possible from the numerical simulations above, because the self-consistent inherently higher SFR of the truncated IMF recipe does not allow the disentanglement of whether the galactic wind is caused by its resulting larger SNeII rate or whether the SNII power is higher due to the later onset of SNeII. 

We consider a star cluster as a single stellar population in the formal mass range of $0.1 - 100\,\msun$ for simplicity. 
Since we are interested in the question of how massive stars affect the energetic regulation by ionizing radiation, stellar winds, and SNII explosions, only the mass range above $8\,\msun$ needs to be specified. We, therefore, split the mass range accordingly: 
\begin{align}
N_{tot} &= k_1\, \int\limits_{0.1\,\msun}^{0.5\,\msun} m^{-1.3}\,dm + k_2 \Big( \int\limits_{0.5\,\msun}^{8\,\msun} m^{-2.3}\, dm\, ,
 + \int\limits_{8\,\msun}^{m_{max}} m^{-2.3}\, dm \Big) \label{eq:N_trunc} \\ 
M_{tot} &= k_3\, \int\limits_{0.1\,\msun}^{0.5\,\msun} m^{-0.3}\,dm + k_4 \Big( \int\limits_{0.5\,\msun}^{8\,\msun} m^{-1.3}\,dm + \int\limits_{8\,\msun}^{m_{max}} m^{-1.3}\,dm \, \Big)\, ,  \label{eq:M_trunc} 
\end{align}
where only the right-hand terms will be evaluated.

\begin{figure}
	\includegraphics[width=1.0\columnwidth]{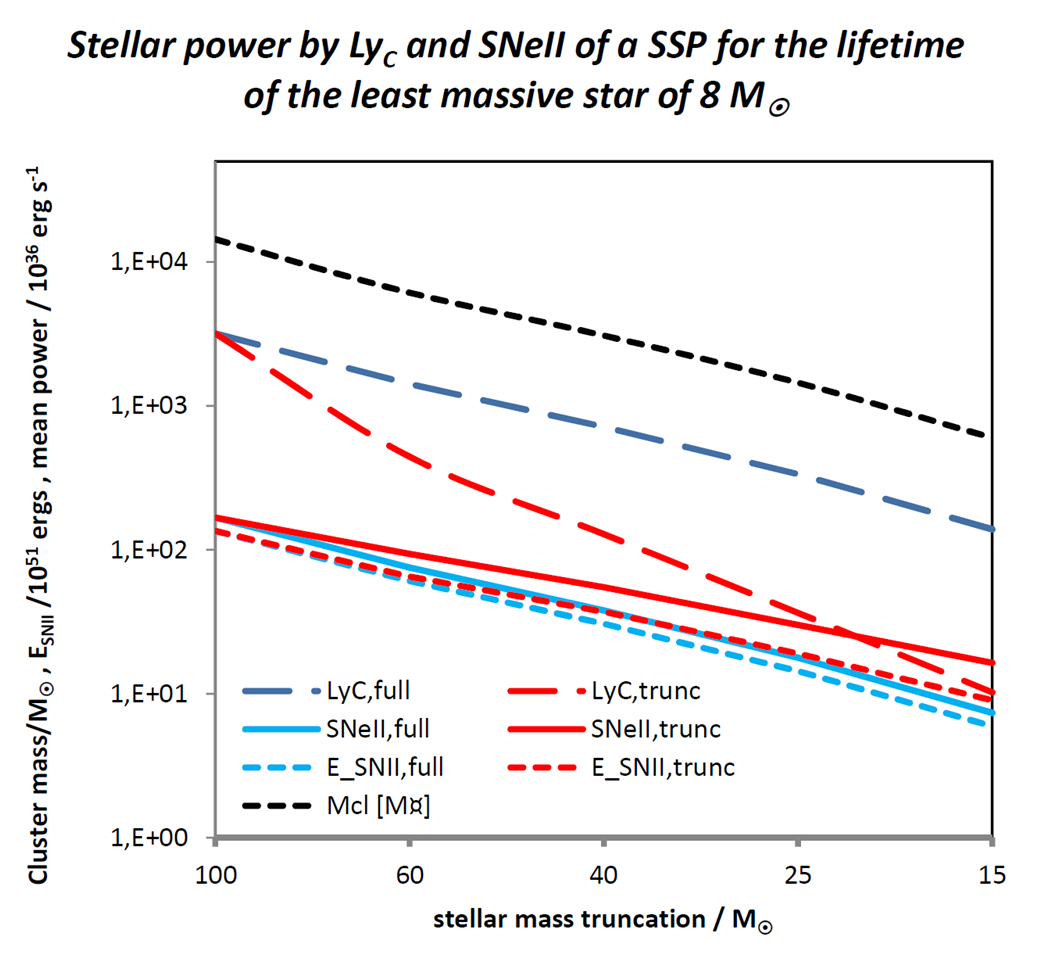}
	\caption{Cluster masses (black short-dashed line), Lyman continuum-radiation power (red/blue long-dashed lines) averaged over the maximum lifetime (of a $8\,\msun$ star), and supernovae type II power (red/blue full-drawn lines, with stellar lifetime delay), and total SNII energies (red/blue short-dashed lines) over the most massive stellar mass $m_{max}$ (values are listed in tab. \ref{tab:imfs}). The power values are normalized to $10^{36}\,\text{erg}\,\text{s}^{-1}$, the total SNII energy release $E_{SNII}$ to $10^{51}\,\text{erg}$, by this, directly counting the number of SNeII.}
	\label{fig:power}
\end{figure}

For a "filled" IMF over the whole mass range up to $m_{max} = 100\,\msun$ the stellar mass fraction contained in the lower mass range of $0.1 - 8\,\msun$ (first two summands in eq. \ref{eq:M_trunc}) amounts to $82.2\,\%$. For a minimally filled IMF, meaning that only a single $100\,\msun$ star exists, and with $30$ logarithmic mass bins dividing the mass range between $8$ and $100\,\msun$, the total cluster mass has to reach about $1.47 \times 10^4\, \msun$. When we reduce this critical maximum mass $m_{max}$ at which a single star is finally formed, lower cluster masses result (see tab. \ref{tab:imfs}). 
We have to emphasize here already, that the results of this analytical study do not depend qualitatively on the number of mass bins as long as the massive-star range is properly sub-divided, i.e. the maximum bin is sufficiently wide enough to include the single stellar mass as demonstrated by \cite{PH2014}. Quantitative uncertainties lie on the percentage level.
For this simple analysis, we first truncate the IMF by fixing the maximum mass $m_{max}$ at 100, 60, 40, 25, and 15 $\msun$ arbitrarily and populate the uppermost mass bin with a single star, so that no star and even no fraction exist above $m_{max}$. 
For this truncated IMF, we calculate the total cluster masses. In the case of the ''filled'' IMF this mass is  distributed accordingly over the whole mass range of $0.1 - 100\,\msun$,  leading to less mass per bin than in the truncated case. Consequently, in the filled IMF even the $m_{max}$ bin and bins above are polulated by number fractions of stars only, which also determine the stellar energy feedback. In fig. \ref{fig:power} the slope of the $m_{max}-log(M_{cl})$ relation of the truncation amounts to $+0.0162$. 

The presented models imply two sources of stellar energy feedback by massive stars: ionizing radiation and SNII explosions, respectively. By the dependence of their power on stellar mass and numbers, differences in the energy release and their effects on the SF self-regulation can be expected and will be explored analytically here. 

Concerning the Lyman continuum radiation, the Ly$_c$ photon flux is a power-law function of the stellar mass $S_{Ly_c}(m) \propto m^{\beta}$ with a large exponent $\beta$ of $+4$ as in eq. \ref{eq:L_ly} or higher (e.g. $+6$ as derived by \citet{hen87}). Folding this $S_{Ly_c}(m)$ mass dependence with the IMF $\xi(m)$ power-law and the (mass-dependent) stellar lifetime with a power of almost $-1.85$, the integration yields the total released $Ly_c$ energy
\begin{equation}
  E_{Ly_c} \propto \int\limits_{8}^{m_{max}}\,S_\ast \xi(m)\, \tau_{\ast}(m)\, dm \propto \int\limits_{8}^{m_{max}}\, m^{0.85}\, dm \propto m_{max}^{1.85} - 46.85 ,
\end{equation} 
leading to a positive power with the maximum mass. 
This also means that the ''filled'' IMF, even with mass fractions 
in the massive-stars' mass bins - in reality impossible, but numerically treatible - produces more $Ly_c$ energy than a truncated one and thus exerts a stronger radiative feedback (see fig. \ref{fig:power}).

\begin{table*}
	\centering
	\caption{Comparison of the cluster mass $M_{cl}$, to form the most massive star with mass $m_{max}$, the SFR to form this cluster mass, the Lyman continuum energy $E_{Ly_c}$, the number of type II SNe $N_{SNeII}$, and the feedback power $L_{Ly_c}$ and $L_{SNeII}$ by Lyman continuum radiation and SNeII, respectively, for filled (full) and truncated (trunc.) IMF. For definition of IMFs see sec. \ref{subsec:imf}. }
\label{tab:imfs}
\begin{tabular}{lccccccc}
\hline\hline
	& $m_{max}\,[\msun]$  & 100 & 60 & 40 & 25 & 15 \\ \hline\hline
  $M_{cl}\,[\msun]$  &  &  14388 & 6120 & 3082 & 1448 & 599 \\ \hline 
  SFR $[\SFR]$ &  &  $1.49 \times 10^{-2}$ & $6.12 \times 10^{-3}$ & $3.08 \times 10^{-3}$ & $1.45 \times 10^{-3}$ & $5.99 \times 10^{-4}$ \\ \hline 
 total mass [$\msun$] & full  & 2619 & 1087 & 547.5 & 257.3 & 106.4 \\ 
$E_{Ly_c}\,[10^{51}\,\text{erg}]$ & full & 2570 & 1170 & 589 & 277 & 114 \\
           & trunc. &  2570 & 357 & 104 & 29.4 & 8.23 \\ \hline
$L_{Ly_c} $       & full & 318 & 143 & 71.5 & 33.6 & 13.9\\
$[10^{37}\,\text{erg}\,\text{s}^{-1}]$ & trunc. & 318 & 44.3 & 12.8 & 3.65 & 1.02 \\ \hline
$N_{SNeII} $ & full & 135.0 & 62.0 & 31.2 & 14.7 & 6.07 \\
           & trunc. &  135 & 65 & 37 & 19 & 9 \\ \hline
$L_{SNeII}$ & full & 16.7 & 7.53 & 3.79 & 1.78 & 0.737 \\
$[10^{37}\,\text{erg}\,\text{s}^{-1}]$ & trunc. &  16.7 & 9.37 & 5.49 & 3.00 & 1.64 \\ \hline
	\hline
\end{tabular}
\end{table*}

The $Ly_c$ radiation acts with a larger power for the filled IMF because the most massive stars exist even with fractions. For the lowest considered truncated cluster mass of 15 $\msun$, this difference amounts to a factor of 13 (tab. \ref{tab:imfs}).  

This situation changes when we consider the SNII energy of both IMF modes. Although the total masses of massive stars only differs insignificantly, i.e. by less than $1\,\%$, between the full and the truncated IMF (tab. \ref{tab:imfs}), the mass in the occupied bins is up to a factor of $2.4$ larger in the truncated IMF. This also holds for the numbers of stars. 
As a side note, it should be mentioned that the star numbers of the truncated IMF are reasonably all set to integer values, while those of the filled IMF have floating values.

The stellar radiative energy feedback by $Ly_c$ photons acts almost continuously during the stellar lifetime, while SNII explosions are instantaneous events for each star, terminating its life. However, for an ensemble of stars with multiple SNeII from the energetic point of view, these events, for simplicity, can be smeared out and treated as a continuous energy source. 
This means, that the sum of SN explosions is taken into account for a time interval from the first occurrence by the shortest-living most massive progenitor star $m_{max}$ until the deaths of the least massive exploding stars, i.e. at $8\,\msun$. Although the total energy of SNeII (see tab. \ref{tab:imfs}) deviates at most for the $m_{max} = 15\,\msun$ model by a factor of $1.5$, their ratio in the SNII power $E_{SNII}$ grows to $2.2$, caused by the shrinkage of the considered time interval from the $m_{max}$ stars to $8\,\msun$. 

As one recognizes from fig. \ref{fig:power}, the $M_{cl}$ curve declines parallelly to the power of SNeII in the full-IMF mode. The ratio of $L_{SNeII}/M_{cl}$ amounts to $3.22\,\text{L}_\odot\,\text{M}_\odot^{-1}$.

\section{Conclusions}\label{sec:conclusion}

In the presented study, we aim at demonstrating by numerical and analytical models, which consequences the neglect of a possible non-standard IMF bear for numerical simulations in cases of low SFRs. 
We apply the simplified approach to fill the IMF from low stellar masses up to a maximum mass at which a full star can be formed. This treatment leads to different star cluster IMFs: truncated IMF when the gas reservoir is consumed and full stellar IMF population otherwise. 
This model with IMF truncation is compared with numerical evolution models of DGs where the IMF is always completely filled even when the IMF allows only fractions of massive stars to be formed. The energetic difference of a filled cluster IMF, despite the concession of star fractions, vs. the truncated IMF, the latter based on complete stars only, are elaborated and strikingly shown: a filled IMF, even by fractions of massive stars, regulates the SF more intensively due to the steep $Ly_c(m)$ power-law than a truncated (or stochastic) IMF. 
This effect can be called SF over-regulations because of the exaggerated energy release. In contrast, the mass fractions of massive stars also contribute only fractions of full SN energy and, by this, does not energize the ISM sufficiently so that cooling at lower temperatures emits the energy too efficiently. It is therefore not sufficient to include SNeII as the only stellar feedback process, because as emphasized the SF self-regulation depends inherently on the $Ly_c$ energetics.\\
\\
The main take-away messages from our models are: At first, although low SFRs e.g. in DGs can produce star clusters which are completely filled by integer stars up to the uppermost mass limit of the IMF, individual star clusters exist which lack a completely filled IMF. 
Secondly, comparing the two extreme cases of filled vs. truncated IMFs leads to the mentioned energetic differences in star-formation self-regulation and SNII energy release. And thirdly, the reduced number of massive stars by a non-standard IMF leaves chemical signatures of enhanced abundance ratios of elements from intermediate-mass to massive stellar progenitors. This latter question and the derivation of SFRs from $\Ha$ luminosities will be addressed in a forthcoming paper.
In order to get a deeper insight into the issues of truncated vs. filled IMF approaches, we consider both for single stellar clusters of various masses analytically. 
The results demonstrate the above-mentioned deviations and agree well with the $M_{cl}-m_{max}$ relations depicted by \cite{WKB10}. We also test the mass distribution in single star clusters by random sampling in order to compare it with the truncation approach. 
We can show that the median of 100 sampling runs for each cluster mass yields an uppermost mass at almost the same range (Hein, BSc thesis, Univ. of Vienna) as the truncation method. From this, one can also conclude that for a given star cluster mass the truncation provides a more easily manageable method than a random sampling procedure.

\section*{Data Availability}
The data underlying this article will be shared on reasonable request to the corresponding author.
The used and described code is available at the Flash Center at the University of Chicago and our extensions are described in section \ref{sec:simulation}.

\section*{Acknowledgements}

The authors benefit from multiple discussions with Pavel Kroupa, Sylvia Ploeckinger, and Simone Recchi, as well as with Janice Lee, Bruce Elmegreen, and Takuji Tsujimoto. 
The authors acknowledge the proof reading by Shelley-Anne Deborah Harrisberg.
We also acknowledge the positive report by an anonymous referee and supportive comments which helped to improve the clarity of the paper. 
The software used in this work was in part developed by the DOE NNSA-ASC OASCR Flash Center at the University of Chicago. The computational results presented here have been achieved using the Vienna Scientific Cluster (VSC) under project no. 70670.



\bibliographystyle{mnras}
\bibliography{papers.bib} 

\begin{thebibliography}{}
\makeatletter
\relax
\def\mn@urlcharsother{\let\do\@makeother \do\$\do\&\do\#\do\^\do\_\do\%\do\~}
\def\mn@doi{\begingroup\mn@urlcharsother \@ifnextchar [ {\mn@doi@}
  {\mn@doi@[]}}
\def\mn@doi@[#1]#2{\def\@tempa{#1}\ifx\@tempa\@empty \href
  {http://dx.doi.org/#2} {doi:#2}\else \href {http://dx.doi.org/#2} {#1}\fi
  \endgroup}
\def\mn@eprint#1#2{\mn@eprint@#1:#2::\@nil}
\def\mn@eprint@arXiv#1{\href {http://arxiv.org/abs/#1} {{\tt arXiv:#1}}}
\def\mn@eprint@dblp#1{\href {http://dblp.uni-trier.de/rec/bibtex/#1.xml}
  {dblp:#1}}
\def\mn@eprint@#1:#2:#3:#4\@nil{\def\@tempa {#1}\def\@tempb {#2}\def\@tempc
  {#3}\ifx \@tempc \@empty \let \@tempc \@tempb \let \@tempb \@tempa \fi \ifx
  \@tempb \@empty \def\@tempb {arXiv}\fi \@ifundefined
  {mn@eprint@\@tempb}{\@tempb:\@tempc}{\expandafter \expandafter \csname
  mn@eprint@\@tempb\endcsname \expandafter{\@tempc}}}

\bibitem[\protect\citeauthoryear{{Agertz}, {Teyssier}  \& {Moore}}{{Agertz}
  et~al.}{2011}]{ATM11}
{Agertz} O.,  {Teyssier} R.,   {Moore} B.,  2011, \mn@doi [\mnras]
  {10.1111/j.1365-2966.2010.17530.x}, \href
  {http://esoads.eso.org/abs/2011MNRAS.410.1391A} {410, 1391}

\bibitem[\protect\citeauthoryear{{Andr\'e}, {Arzoumanian, D.}, {K\"onyves, V.},
  {Shimajiri, Y.}  \& {Palmeirim, P.}}{{Andr\'e} et~al.}{2019}]{And19}
{Andr\'e} P.,  {Arzoumanian, D.} {K\"onyves, V.} {Shimajiri, Y.}  {Palmeirim,
  P.} 2019, \mn@doi [A\&A] {10.1051/0004-6361/201935915}, 629, L4

\bibitem[\protect\citeauthoryear{{Andrews} et~al.,}{{Andrews}
  et~al.}{2013}]{ACC13}
{Andrews} J.~E.,  et~al., 2013, \mn@doi [\apj] {10.1088/0004-637X/767/1/51},
  \href {http://esoads.eso.org/abs/2013ApJ...767...51A} {767, 51}

\bibitem[\protect\citeauthoryear{{Andrews} et~al.,}{{Andrews}
  et~al.}{2014}]{ACC14}
{Andrews} J.~E.,  et~al., 2014, \mn@doi [\apj] {10.1088/0004-637X/793/1/4},
  \href {http://esoads.eso.org/abs/2014ApJ...793....4A} {793, 4}

\bibitem[\protect\citeauthoryear{{Applebaum}, {Brooks}, {Quinn}  \&
  {Christensen}}{{Applebaum} et~al.}{2020}]{App20}
{Applebaum} E.,  {Brooks} A.~M.,  {Quinn} T.~R.,   {Christensen} C.~R.,  2020,
  \mn@doi [\mnras] {10.1093/mnras/stz3331}, \href
  {https://ui.adsabs.harvard.edu/abs/2020MNRAS.492....8A} {492, 8}

\bibitem[\protect\citeauthoryear{{Baugh}, {Lacey}, {Frenk}, {Granato}, {Silva},
  {Bressan}, {Benson}  \& {Cole}}{{Baugh} et~al.}{2005}]{BLF05}
{Baugh} C.~M.,  {Lacey} C.~G.,  {Frenk} C.~S.,  {Granato} G.~L.,  {Silva} L.,
  {Bressan} A.,  {Benson} A.~J.,   {Cole} S.,  2005, \mn@doi [\mnras]
  {10.1111/j.1365-2966.2004.08553.x}, \href
  {https://ui.adsabs.harvard.edu/abs/2005MNRAS.356.1191B} {356, 1191}

\bibitem[\protect\citeauthoryear{Baumschlager, Hensler, Steyrleithner  \&
  Recchi}{Baumschlager et~al.}{2018}]{Baum19}
Baumschlager B.,  Hensler G.,  Steyrleithner P.,   Recchi S.,  2018, Monthly
  Notices of the Royal Astronomical Society, 483, 5315

\bibitem[\protect\citeauthoryear{{Bekki}}{{Bekki}}{2013}]{Bekki13}
{Bekki} K.,  2013, \mn@doi [\mnras] {10.1093/mnras/stt1735}, \href
  {http://esoads.eso.org/abs/2013MNRAS.436.2254B} {436, 2254}

\bibitem[\protect\citeauthoryear{{Boehringer} \& {Hensler}}{{Boehringer} \&
  {Hensler}}{1989}]{BH89}
{Boehringer} H.,  {Hensler} G.,  1989, \aap, \href
  {http://esoads.eso.org/abs/1989A%26A...215..147B} {215, 147}

\bibitem[\protect\citeauthoryear{{Bonnell}, {Bate}  \& {Vine}}{{Bonnell}
  et~al.}{2003}]{BBV03}
{Bonnell} I.~A.,  {Bate} M.~R.,   {Vine} S.~G.,  2003, \mn@doi [\mnras]
  {10.1046/j.1365-8711.2003.06687.x}, \href
  {https://ui.adsabs.harvard.edu/abs/2003MNRAS.343..413B} {343, 413}

\bibitem[\protect\citeauthoryear{{Boselli}, {Boissier}, {Cortese}, {Buat},
  {Hughes}  \& {Gavazzi}}{{Boselli} et~al.}{2009}]{Bos09}
{Boselli} A.,  {Boissier} S.,  {Cortese} L.,  {Buat} V.,  {Hughes} T.~M.,
  {Gavazzi} G.,  2009, \mn@doi [\apj] {10.1088/0004-637X/706/2/1527}, \href
  {https://ui.adsabs.harvard.edu/abs/2009ApJ...706.1527B} {706, 1527}

\bibitem[\protect\citeauthoryear{{Boselli} et~al.,}{{Boselli}
  et~al.}{2018}]{Bos18}
{Boselli} A.,  et~al., 2018, \mn@doi [\aap] {10.1051/0004-6361/201732410},
  \href {https://ui.adsabs.harvard.edu/abs/2018A&A...615A.114B} {615, A114}

\bibitem[\protect\citeauthoryear{{Calzetti}, {Chandar}, {Lee}, {Elmegreen},
  {Kennicutt}  \& {Whitmore}}{{Calzetti} et~al.}{2010}]{Cal10}
{Calzetti} D.,  {Chandar} R.,  {Lee} J.~C.,  {Elmegreen} B.~G.,  {Kennicutt}
  R.~C.,   {Whitmore} B.,  2010, \mn@doi [\apjl]
  {10.1088/2041-8205/719/2/L158}, \href
  {https://ui.adsabs.harvard.edu/abs/2010ApJ...719L.158C} {719, L158}

\bibitem[\protect\citeauthoryear{{Chabrier}}{{Chabrier}}{2003}]{Cha03}
{Chabrier} G.,  2003, \mn@doi [\pasp] {10.1086/376392}, \href
  {https://ui.adsabs.harvard.edu/abs/2003PASP..115..763C} {115, 763}

\bibitem[\protect\citeauthoryear{{Chamandy} \& {Shukurov}}{{Chamandy} \&
  {Shukurov}}{2020}]{CS20}
{Chamandy} L.,  {Shukurov} A.,  2020, \mn@doi [Galaxies]
  {10.3390/galaxies8030056}, \href
  {https://ui.adsabs.harvard.edu/abs/2020Galax...8...56C} {8, 56}

\bibitem[\protect\citeauthoryear{{Conroy} \& {van Dokkum}}{{Conroy} \& {van
  Dokkum}}{2012}]{CvD12}
{Conroy} C.,  {van Dokkum} P.,  2012, \mn@doi [\apj]
  {10.1088/0004-637X/747/1/69}, \href
  {https://ui.adsabs.harvard.edu/abs/2012ApJ...747...69C} {747, 69}

\bibitem[\protect\citeauthoryear{{Dalgarno} \& {McCray}}{{Dalgarno} \&
  {McCray}}{1972}]{DM72}
{Dalgarno} A.,  {McCray} R.~A.,  1972, \mn@doi [\araa]
  {10.1146/annurev.aa.10.090172.002111}, \href
  {http://esoads.eso.org/abs/1972ARA%26A..10..375D} {10, 375}

\bibitem[\protect\citeauthoryear{{Dib}, {Schmeja}  \& {Hony}}{{Dib}
  et~al.}{2017}]{DSH17}
{Dib} S.,  {Schmeja} S.,   {Hony} S.,  2017, \mn@doi [\mnras]
  {10.1093/mnras/stw2465}, \href
  {https://ui.adsabs.harvard.edu/abs/2017MNRAS.464.1738D} {464, 1738}

\bibitem[\protect\citeauthoryear{{Eldridge}}{{Eldridge}}{2012}]{Eldridge12}
{Eldridge} J.~J.,  2012, \mn@doi [\mnras] {10.1111/j.1365-2966.2012.20662.x},
  \href {https://ui.adsabs.harvard.edu/abs/2012MNRAS.422..794E} {422, 794}

\bibitem[\protect\citeauthoryear{{Elmegreen}}{{Elmegreen}}{2000}]{Elmegreen2000}
{Elmegreen} B.~G.,  2000, \mn@doi [\apj] {10.1086/309204}, \href
  {https://ui.adsabs.harvard.edu/abs/2000ApJ...539..342E} {539, 342}

\bibitem[\protect\citeauthoryear{Elmegreen}{Elmegreen}{2006}]{Elm06}
Elmegreen B.~G.,  2006, \mn@doi [The Astrophysical Journal] {10.1086/505785},
  648, 572

\bibitem[\protect\citeauthoryear{{Elmegreen}}{{Elmegreen}}{2009}]{Elme09}
{Elmegreen} B.~G.,  2009, in {Sheth} K.,  {Noriega-Crespo} A.,  {Ingalls}
  J.~G.,   {Paladini} R.,  eds, The Evolving ISM in the Milky Way and Nearby
  Galaxies. p.~14 (\mn@eprint {arXiv} {0803.3154})

\bibitem[\protect\citeauthoryear{Elmegreen et~al.,}{Elmegreen
  et~al.}{2014}]{EE14}
Elmegreen D.~M.,  et~al., 2014, \mn@doi [The Astrophysical Journal]
  {10.1088/2041-8205/787/1/l15}, 787, L15

\bibitem[\protect\citeauthoryear{{Faesi}, {Lada}  \& {Forbrich}}{{Faesi}
  et~al.}{2019}]{Faesi19}
{Faesi} C.~M.,  {Lada} C.~J.,   {Forbrich} J.,  2019, VizieR Online Data
  Catalog, \href {https://ui.adsabs.harvard.edu/abs/2019yCat..18570019F} {p.
  J/ApJ/857/19}

\bibitem[\protect\citeauthoryear{{Ferreras}, {La Barbera}, {de La Rosa},
  {Vazdekis}, {de Carvalho}, {Falcon-Barroso}  \& {Ricciardelli}}{{Ferreras}
  et~al.}{2013}]{Fer13}
{Ferreras} I.,  {La Barbera} F.,  {de La Rosa} I.~G.,  {Vazdekis} A.,  {de
  Carvalho} R.~R.,  {Falcon-Barroso} J.,   {Ricciardelli} E.,  2013, \mn@doi
  [\mnras] {10.1093/mnrasl/sls014}, \href
  {https://ui.adsabs.harvard.edu/abs/2013MNRAS.429L..15F} {429, L15}

\bibitem[\protect\citeauthoryear{{Freyer}, {Hensler}  \& {Yorke}}{{Freyer}
  et~al.}{2003}]{FHY03}
{Freyer} T.,  {Hensler} G.,   {Yorke} H.~W.,  2003, \mn@doi [\apj]
  {10.1086/376937}, \href
  {https://ui.adsabs.harvard.edu/abs/2003ApJ...594..888F} {594, 888}

\bibitem[\protect\citeauthoryear{{Freyer}, {Hensler}  \& {Yorke}}{{Freyer}
  et~al.}{2006}]{FHY06}
{Freyer} T.,  {Hensler} G.,   {Yorke} H.~W.,  2006, \mn@doi [\apj]
  {10.1086/498734}, \href
  {https://ui.adsabs.harvard.edu/abs/2006ApJ...638..262F} {638, 262}

\bibitem[\protect\citeauthoryear{{Fryxell} et~al.,}{{Fryxell}
  et~al.}{2000}]{Fryxell00}
{Fryxell} B.,  et~al., 2000, \mn@doi [\apjs] {10.1086/317361}, \href
  {http://esoads.eso.org/abs/2000ApJS..131..273F} {131, 273}

\bibitem[\protect\citeauthoryear{{Fumagalli}, {Gavazzi}, {Scaramella}  \&
  {Franzetti}}{{Fumagalli} et~al.}{2011}]{FGS11}
{Fumagalli} M.,  {Gavazzi} G.,  {Scaramella} R.,   {Franzetti} P.,  2011, AA,
  528, A46

\bibitem[\protect\citeauthoryear{{Gatto} et~al.,}{{Gatto}
  et~al.}{2017}]{Gatto17}
{Gatto} A.,  et~al., 2017, \mn@doi [\mnras] {10.1093/mnras/stw3209}, \href
  {https://ui.adsabs.harvard.edu/abs/2017MNRAS.466.1903G} {466, 1903}

\bibitem[\protect\citeauthoryear{{Governato} et~al.,}{{Governato}
  et~al.}{2012}]{Gov12}
{Governato} F.,  et~al., 2012, \mn@doi [\mnras]
  {10.1111/j.1365-2966.2012.20696.x}, \href
  {https://ui.adsabs.harvard.edu/abs/2012MNRAS.422.1231G} {422, 1231}

\bibitem[\protect\citeauthoryear{{Grasha} et~al.,}{{Grasha}
  et~al.}{2019}]{Grasha19}
{Grasha} K.,  et~al., 2019, \mn@doi [\mnras] {10.1093/mnras/sty3424}, \href
  {https://ui.adsabs.harvard.edu/abs/2019MNRAS.483.4707G} {483, 4707}

\bibitem[\protect\citeauthoryear{{Greggio}}{{Greggio}}{2010}]{Greggio10}
{Greggio} L.,  2010, \mn@doi [\mnras] {10.1111/j.1365-2966.2010.16371.x}, \href
  {https://ui.adsabs.harvard.edu/abs/2010MNRAS.406...22G} {406, 22}

\bibitem[\protect\citeauthoryear{{Hensler}}{{Hensler}}{1987}]{hen87}
{Hensler} G.,  1987, Mitteilungen der Astronomischen Gesellschaft Hamburg,
  \href {https://ui.adsabs.harvard.edu/abs/1987MitAG..70..141H} {70, 141}

\bibitem[\protect\citeauthoryear{{Hensler}}{{Hensler}}{2007}]{Hen07}
{Hensler} G.,  2007, in {Emsellem} E.,  {Wozniak} H.,  {Massacrier} G.,
  {Gonzalez} J.~F.,  {Devriendt} J.,   {Champavert} N.,  eds,  EAS Publications
  Series Vol. 24, EAS Publications Series. pp 113--118 (\mn@eprint {arXiv}
  {0709.0631}), \mn@doi{10.1051/eas:2007018}

\bibitem[\protect\citeauthoryear{{Hensler}, {Theis}  \& {Gallagher}}{{Hensler}
  et~al.}{2004}]{HTG04}
{Hensler} G.,  {Theis} C.,   {Gallagher} III. J.~S.,  2004, \mn@doi [\aap]
  {10.1051/0004-6361:20048002}, \href
  {http://esoads.eso.org/abs/2004A%26A...426...25H} {426, 25}

\bibitem[\protect\citeauthoryear{{Hester} et~al.,}{{Hester}
  et~al.}{2010}]{HSN10}
{Hester} J.~A.,  et~al., 2010, ApJL, 716, L14

\bibitem[\protect\citeauthoryear{{Hunter}, {Elmegreen}, {Dupuy}  \&
  {Mortonson}}{{Hunter} et~al.}{2003}]{HED03}
{Hunter} D.~A.,  {Elmegreen} B.~G.,  {Dupuy} T.~J.,   {Mortonson} M.,  2003,
  \mn@doi [\aj] {10.1086/378056}, \href
  {https://ui.adsabs.harvard.edu/abs/2003AJ....126.1836H} {126, 1836}

\bibitem[\protect\citeauthoryear{{J{\'a}chym}, {Combes}, {Cortese}, {Sun}  \&
  {Kenney}}{{J{\'a}chym} et~al.}{2014}]{JCC14}
{J{\'a}chym} P.,  {Combes} F.,  {Cortese} L.,  {Sun} M.,   {Kenney} J.~D.~P.,
  2014, ApJ, 792, 11

\bibitem[\protect\citeauthoryear{{Jeffreson}, {Krumholz}, {Fujimoto},
  {Armillotta}, {Keller}, {Chevance}  \& {Kruijssen}}{{Jeffreson}
  et~al.}{2021}]{JKF21}
{Jeffreson} S. M.~R.,  {Krumholz} M.~R.,  {Fujimoto} Y.,  {Armillotta} L.,
  {Keller} B.~W.,  {Chevance} M.,   {Kruijssen} J.~M.~D.,  2021, \mn@doi
  [\mnras] {10.1093/mnras/stab1536}, \href
  {https://ui.adsabs.harvard.edu/abs/2021MNRAS.505.3470J} {505, 3470}

\bibitem[\protect\citeauthoryear{{Johnson} et~al.,}{{Johnson}
  et~al.}{2015}]{JKK15}
{Johnson} M.~C.,  et~al., 2015, MNRAS, 451, 3192

\bibitem[\protect\citeauthoryear{{Johnson} et~al.,}{{Johnson}
  et~al.}{2017}]{JSD17}
{Johnson} L.~C.,  et~al., 2017, \mn@doi [\apj] {10.3847/1538-4357/aa6a1f},
  \href {https://ui.adsabs.harvard.edu/abs/2017ApJ...839...78J} {839, 78}

\bibitem[\protect\citeauthoryear{{Kenney}, {Geha}, {J{\'a}chym}, {Crowl},
  {Dague}, {Chung}, {van Gorkom}  \& {Vollmer}}{{Kenney} et~al.}{2014}]{KGJ14}
{Kenney} J.~D.~P.,  {Geha} M.,  {J{\'a}chym} P.,  {Crowl} H.~H.,  {Dague} W.,
  {Chung} A.,  {van Gorkom} J.,   {Vollmer} B.,  2014, ApJ, 780, 119

\bibitem[\protect\citeauthoryear{{Kennicutt}}{{Kennicutt}}{1998}]{Ken98}
{Kennicutt} Robert~C. J.,  1998, \mn@doi [\apj] {10.1086/305588}, \href
  {https://ui.adsabs.harvard.edu/abs/1998ApJ...498..541K} {498, 541}

\bibitem[\protect\citeauthoryear{{Kirby}, {Cohen}, {Guhathakurta}, {Cheng},
  {Bullock}  \& {Gallazzi}}{{Kirby} et~al.}{2013}]{Kirby13}
{Kirby} E.~N.,  {Cohen} J.~G.,  {Guhathakurta} P.,  {Cheng} L.,  {Bullock}
  J.~S.,   {Gallazzi} A.,  2013, \mn@doi [\apj] {10.1088/0004-637X/779/2/102},
  \href {https://ui.adsabs.harvard.edu/abs/2013ApJ...779..102K} {779, 102}

\bibitem[\protect\citeauthoryear{{Koeppen}, {Theis}  \& {Hensler}}{{Koeppen}
  et~al.}{1995}]{KTH95}
{Koeppen} J.,  {Theis} C.,   {Hensler} G.,  1995, \aap, \href
  {http://esoads.eso.org/abs/1995A%26A...296...99K} {296, 99}

\bibitem[\protect\citeauthoryear{{Kroupa}}{{Kroupa}}{2001}]{Kroupa01}
{Kroupa} P.,  2001, \mn@doi [\mnras] {10.1046/j.1365-8711.2001.04022.x}, \href
  {http://esoads.eso.org/abs/2001MNRAS.322..231K} {322, 231}

\bibitem[\protect\citeauthoryear{{Kroupa}}{{Kroupa}}{2002}]{Kroupa02}
{Kroupa} P.,  2002, \mn@doi [Science] {10.1126/science.1067524}, \href
  {http://esoads.eso.org/abs/2002Sci...295...82K} {295, 82}

\bibitem[\protect\citeauthoryear{{Kroupa}}{{Kroupa}}{2014}]{KroupaRev14}
{Kroupa} P.,  2014, \mn@doi [Astrophysics and Space Science Proceedings]
  {10.1007/978-3-319-03041-8_65}, \href
  {http://esoads.eso.org/abs/2014ASSP...36..335K} {36, 335}

\bibitem[\protect\citeauthoryear{{Kroupa}, {Weidner}, {Pflamm-Altenburg},
  {Thies}, {Dabringhausen}, {Marks}  \& {Maschberger}}{{Kroupa}
  et~al.}{2013}]{KWP13}
{Kroupa} P.,  {Weidner} C.,  {Pflamm-Altenburg} J.,  {Thies} I.,
  {Dabringhausen} J.,  {Marks} M.,   {Maschberger} T.,  2013, {The Stellar and
  Sub-Stellar Initial Mass Function of Simple and Composite Populations}.
p.~115, \mn@doi{10.1007/978-94-007-5612-0_4}

\bibitem[\protect\citeauthoryear{{Krumholz}}{{Krumholz}}{2014}]{Krumholz14}
{Krumholz} M.~R.,  2014, \mn@doi [\physrep] {10.1016/j.physrep.2014.02.001},
  \href {http://esoads.eso.org/abs/2014PhR...539...49K} {539, 49}

\bibitem[\protect\citeauthoryear{{Krumholz}, {McKee}  \&
  {Bland-Hawthorn}}{{Krumholz} et~al.}{2019}]{Krum19}
{Krumholz} M.~R.,  {McKee} C.~F.,   {Bland-Hawthorn} J.,  2019, \mn@doi [\araa]
  {10.1146/annurev-astro-091918-104430}, \href
  {https://ui.adsabs.harvard.edu/abs/2019ARA&A..57..227K} {57, 227}

\bibitem[\protect\citeauthoryear{{Kudritzki}, {Pauldrach}  \&
  {Puls}}{{Kudritzki} et~al.}{1987}]{KPP87}
{Kudritzki} R.~P.,  {Pauldrach} A.,   {Puls} J.,  1987, \aap, \href
  {https://ui.adsabs.harvard.edu/abs/1987A&A...173..293K} {173, 293}

\bibitem[\protect\citeauthoryear{{Lada} \& {Lada}}{{Lada} \&
  {Lada}}{2003}]{LadaLada03}
{Lada} C.~J.,  {Lada} E.~A.,  2003, \mn@doi [\araa]
  {10.1146/annurev.astro.41.011802.094844}, \href
  {https://ui.adsabs.harvard.edu/abs/2003ARA&A..41...57L} {41, 57}

\bibitem[\protect\citeauthoryear{{Larsen}}{{Larsen}}{2002}]{Lar02}
{Larsen} S.~S.,  2002, \mn@doi [\aj] {10.1086/342381}, \href
  {https://ui.adsabs.harvard.edu/abs/2002AJ....124.1393L} {124, 1393}

\bibitem[\protect\citeauthoryear{{Larsen}}{{Larsen}}{2009}]{Larsen09}
{Larsen} S.~S.,  2009, \mn@doi [\aap] {10.1051/0004-6361:200811212}, \href
  {https://ui.adsabs.harvard.edu/abs/2009A&A...494..539L} {494, 539}

\bibitem[\protect\citeauthoryear{{Larson}}{{Larson}}{1988}]{Larson88}
{Larson} R.~B.,  1988, in {Pudritz} R.~E.,  {Fich} M.,  eds,  NATO Advanced
  Study Institute (ASI) Series C Vol. 232, Galactic and Extragalactic Star
  Formation. p.~459, \mn@doi{10.1007/978-94-009-2973-9_27}

\bibitem[\protect\citeauthoryear{{Lee-Waddell} et~al.,}{{Lee-Waddell}
  et~al.}{2016}]{LeeW16}
{Lee-Waddell} K.,  et~al., 2016, \mn@doi [\mnras] {10.1093/mnras/stw1162},
  \href {https://ui.adsabs.harvard.edu/abs/2016MNRAS.460.2945L} {460, 2945}

\bibitem[\protect\citeauthoryear{{Lee-Waddell}, {Madrid}, {Spekkens},
  {Donzelli}, {Koribalski}, {Serra}  \& {Cannon}}{{Lee-Waddell}
  et~al.}{2018}]{LeeW18}
{Lee-Waddell} K.,  {Madrid} J.~P.,  {Spekkens} K.,  {Donzelli} C.~J.,
  {Koribalski} B.~S.,  {Serra} P.,   {Cannon} J.,  2018, \mn@doi [\mnras]
  {10.1093/mnras/sty2042}, \href
  {https://ui.adsabs.harvard.edu/abs/2018MNRAS.480.2719L} {480, 2719}

\bibitem[\protect\citeauthoryear{{Lee}, {Gibson}, {Flynn}, {Kawata}  \&
  {Beasley}}{{Lee} et~al.}{2004}]{Lee04}
{Lee} H.-c.,  {Gibson} B.~K.,  {Flynn} C.,  {Kawata} D.,   {Beasley} M.~A.,
  2004, \mn@doi [\mnras] {10.1111/j.1365-2966.2004.08049.x}, \href
  {https://ui.adsabs.harvard.edu/abs/2004MNRAS.353..113L} {353, 113}

\bibitem[\protect\citeauthoryear{{Lee} et~al.,}{{Lee} et~al.}{2009}]{Lee09}
{Lee} J.~C.,  et~al., 2009, \mn@doi [\apj] {10.1088/0004-637X/706/1/599}, \href
  {http://esoads.eso.org/abs/2009ApJ...706..599L} {706, 599}

\bibitem[\protect\citeauthoryear{{Lombardi}, {Alves}  \& {Lada}}{{Lombardi}
  et~al.}{2015}]{LAL15}
{Lombardi} M.,  {Alves} J.,   {Lada} C.~J.,  2015, \mn@doi [\aap]
  {10.1051/0004-6361/201525650}, \href
  {https://ui.adsabs.harvard.edu/abs/2015A&A...576L...1L} {576, L1}

\bibitem[\protect\citeauthoryear{Loubser, Hoekstra, Babul, Bahé  \&
  Donahue}{Loubser et~al.}{2020}]{Lou21}
Loubser S.~I.,  Hoekstra H.,  Babul A.,  Bahé Y.~M.,   Donahue M.,  2020,
  Monthly Notices of the Royal Astronomical Society, 500, 4153

\bibitem[\protect\citeauthoryear{{Maeder}}{{Maeder}}{1996}]{Maeder96}
{Maeder} A.,  1996, in {Leitherer} C.,  {Fritze-von-Alvensleben} U.,   {Huchra}
  J.,  eds,  Astronomical Society of the Pacific Conference Series Vol. 98,
  From Stars to Galaxies: the Impact of Stellar Physics on Galaxy Evolution.
  p.~141

\bibitem[\protect\citeauthoryear{{Maoz}, {Mannucci}  \& {Brandt}}{{Maoz}
  et~al.}{2012}]{MMB12}
{Maoz} D.,  {Mannucci} F.,   {Brandt} T.~D.,  2012, \mn@doi [\mnras]
  {10.1111/j.1365-2966.2012.21871.x}, \href
  {https://ui.adsabs.harvard.edu/abs/2012MNRAS.426.3282M} {426, 3282}

\bibitem[\protect\citeauthoryear{{Marigo}, {Bressan}  \& {Chiosi}}{{Marigo}
  et~al.}{1996}]{MBC96}
{Marigo} P.,  {Bressan} A.,   {Chiosi} C.,  1996, \aap, \href
  {http://esoads.eso.org/abs/1996A%26A...313..545M} {313, 545}

\bibitem[\protect\citeauthoryear{{Mart{\'\i}n-Navarro}
  et~al.,}{{Mart{\'\i}n-Navarro} et~al.}{2015}]{MaNa15}
{Mart{\'\i}n-Navarro} I.,  et~al., 2015, \mn@doi [\apjl]
  {10.1088/2041-8205/806/2/L31}, \href
  {https://ui.adsabs.harvard.edu/abs/2015ApJ...806L..31M} {806, L31}

\bibitem[\protect\citeauthoryear{McWilliam, Wallerstein  \& Mottini}{McWilliam
  et~al.}{2013}]{McW13}
McWilliam A.,  Wallerstein G.,   Mottini M.,  2013, \mn@doi [The Astrophysical
  Journal] {10.1088/0004-637x/778/2/149}, 778, 149

\bibitem[\protect\citeauthoryear{{Melekh}, {Recchi}, {Hensler}  \&
  {Buhajenko}}{{Melekh} et~al.}{2015}]{MRH15}
{Melekh} B.,  {Recchi} S.,  {Hensler} G.,   {Buhajenko} O.,  2015, \mn@doi
  [\mnras] {10.1093/mnras/stv569}, \href
  {http://esoads.eso.org/abs/2015MNRAS.450..111M} {450, 111}

\bibitem[\protect\citeauthoryear{{Meurer} et~al.,}{{Meurer}
  et~al.}{2009}]{Meu09}
{Meurer} G.~R.,  et~al., 2009, \mn@doi [\apj] {10.1088/0004-637X/695/1/765},
  \href {https://ui.adsabs.harvard.edu/abs/2009ApJ...695..765M} {695, 765}

\bibitem[\protect\citeauthoryear{{Nagashima}, {Lacey}, {Okamoto}, {Baugh},
  {Frenk}  \& {Cole}}{{Nagashima} et~al.}{2005}]{NLO05}
{Nagashima} M.,  {Lacey} C.~G.,  {Okamoto} T.,  {Baugh} C.~M.,  {Frenk} C.~S.,
   {Cole} S.,  2005, \mn@doi [\mnras] {10.1111/j.1745-3933.2005.00078.x}, \href
  {https://ui.adsabs.harvard.edu/abs/2005MNRAS.363L..31N} {363, L31}

\bibitem[\protect\citeauthoryear{Pelkonen, Padoan, Haugb{\o}lle  \&
  Nordlund}{Pelkonen et~al.}{2021}]{Pel21}
Pelkonen V.-M.,  Padoan P.,  Haugb{\o}lle T.,   Nordlund {\AA}.,  2021, \mn@doi
  [Monthly Notices of the Royal Astronomical Society] {10.1093/mnras/stab844},
  504, 1219

\bibitem[\protect\citeauthoryear{{Pflamm-Altenburg}, {Weidner}  \&
  {Kroupa}}{{Pflamm-Altenburg} et~al.}{2007}]{PWK07}
{Pflamm-Altenburg} J.,  {Weidner} C.,   {Kroupa} P.,  2007, \mn@doi [\apj]
  {10.1086/523033}, \href {http://esoads.eso.org/abs/2007ApJ...671.1550P} {671,
  1550}

\bibitem[\protect\citeauthoryear{{Ploeckinger}, {Hensler}, {Recchi}, {Mitchell}
   \& {Kroupa}}{{Ploeckinger} et~al.}{2014}]{PHR14}
{Ploeckinger} S.,  {Hensler} G.,  {Recchi} S.,  {Mitchell} N.,   {Kroupa} P.,
  2014, \mn@doi [\mnras] {10.1093/mnras/stt2211}, \href
  {http://esoads.eso.org/abs/2014MNRAS.437.3980P} {437, 3980}

\bibitem[\protect\citeauthoryear{{Ploeckinger}, {Recchi}, {Hensler}  \&
  {Kroupa}}{{Ploeckinger} et~al.}{2015}]{PRH15}
{Ploeckinger} S.,  {Recchi} S.,  {Hensler} G.,   {Kroupa} P.,  2015, \mn@doi
  [\mnras] {10.1093/mnras/stu2629}, \href
  {http://esoads.eso.org/abs/2015MNRAS.447.2512P} {447, 2512}

\bibitem[\protect\citeauthoryear{{Popescu} \& {Hanson}}{{Popescu} \&
  {Hanson}}{2014}]{PH2014}
{Popescu} B.,  {Hanson} M.~M.,  2014, \mn@doi [\apj]
  {10.1088/0004-637X/780/1/27}, \href
  {https://ui.adsabs.harvard.edu/abs/2014ApJ...780...27P} {780, 27}

\bibitem[\protect\citeauthoryear{{Portinari}, {Chiosi}  \&
  {Bressan}}{{Portinari} et~al.}{1998}]{PCB98}
{Portinari} L.,  {Chiosi} C.,   {Bressan} A.,  1998, \aap, \href
  {http://esoads.eso.org/abs/1998A%26A...334..505P} {334, 505}

\bibitem[\protect\citeauthoryear{{Recchi} \& {Hensler}}{{Recchi} \&
  {Hensler}}{2013}]{RH13}
{Recchi} S.,  {Hensler} G.,  2013, \mn@doi [\aap]
  {10.1051/0004-6361/201220845}, \href
  {http://esoads.eso.org/abs/2013A%26A...551A..41R} {551, A41}

\bibitem[\protect\citeauthoryear{{Recchi}, {Spitoni}, {Matteucci}  \&
  {Lanfranchi}}{{Recchi} et~al.}{2008}]{Rec08}
{Recchi} S.,  {Spitoni} E.,  {Matteucci} F.,   {Lanfranchi} G.~A.,  2008,
  \mn@doi [\aap] {10.1051/0004-6361:200809879}, \href
  {https://ui.adsabs.harvard.edu/abs/2008A&A...489..555R} {489, 555}

\bibitem[\protect\citeauthoryear{{Recchi}, {Calura}  \& {Kroupa}}{{Recchi}
  et~al.}{2009}]{RCK09}
{Recchi} S.,  {Calura} F.,   {Kroupa} P.,  2009, \mn@doi [\aap]
  {10.1051/0004-6361/200811472}, \href
  {http://esoads.eso.org/abs/2009A%26A...499..711R} {499, 711}

\bibitem[\protect\citeauthoryear{{Roychowdhury}, {Chengalur}, {Begum}  \&
  {Karachentsev}}{{Roychowdhury} et~al.}{2009}]{Roy09}
{Roychowdhury} S.,  {Chengalur} J.~N.,  {Begum} A.,   {Karachentsev} I.~D.,
  2009, \mn@doi [\mnras] {10.1111/j.1365-2966.2009.14931.x}, \href
  {https://ui.adsabs.harvard.edu/abs/2009MNRAS.397.1435R} {397, 1435}

\bibitem[\protect\citeauthoryear{{Salpeter}}{{Salpeter}}{1955}]{Salpeter55}
{Salpeter} E.~E.,  1955, \mn@doi [\apj] {10.1086/145971}, \href
  {http://esoads.eso.org/abs/1955ApJ...121..161S} {121, 161}

\bibitem[\protect\citeauthoryear{{Schroyen}, {de Rijcke}, {Valcke},
  {Cloet-Osselaer}  \& {Dejonghe}}{{Schroyen} et~al.}{2011}]{Schroy11}
{Schroyen} J.,  {de Rijcke} S.,  {Valcke} S.,  {Cloet-Osselaer} A.,
  {Dejonghe} H.,  2011, \mn@doi [\mnras] {10.1111/j.1365-2966.2011.19083.x},
  \href {https://ui.adsabs.harvard.edu/abs/2011MNRAS.416..601S} {416, 601}

\bibitem[\protect\citeauthoryear{{Schure}, {Kosenko}, {Kaastra}, {Keppens}  \&
  {Vink}}{{Schure} et~al.}{2009}]{SKK09}
{Schure} K.~M.,  {Kosenko} D.,  {Kaastra} J.~S.,  {Keppens} R.,   {Vink} J.,
  2009, \mn@doi [\aap] {10.1051/0004-6361/200912495}, \href
  {http://esoads.eso.org/abs/2009A%26A...508..751S} {508, 751}

\bibitem[\protect\citeauthoryear{{Smith}}{{Smith}}{2020}]{Smith20}
{Smith} R.~J.,  2020, \mn@doi [\araa] {10.1146/annurev-astro-032620-020217},
  \href {https://ui.adsabs.harvard.edu/abs/2020ARA&A..58..577S} {58, 577}

\bibitem[\protect\citeauthoryear{{Smith}}{{Smith}}{2021}]{Smith21}
{Smith} M.~C.,  2021, \mn@doi [\mnras] {10.1093/mnras/stab291}, \href
  {https://ui.adsabs.harvard.edu/abs/2021MNRAS.502.5417S} {502, 5417}

\bibitem[\protect\citeauthoryear{Steyrleithner, Hensler  \&
  Boselli}{Steyrleithner et~al.}{2020}]{Steyr20}
Steyrleithner P.,  Hensler G.,   Boselli A.,  2020, Monthly Notices of the
  Royal Astronomical Society, 494, 1114

\bibitem[\protect\citeauthoryear{{Taylor} et~al.,}{{Taylor}
  et~al.}{2011}]{GAMA11}
{Taylor} E.~N.,  et~al., 2011, \mn@doi [\mnras]
  {10.1111/j.1365-2966.2011.19536.x}, \href
  {https://ui.adsabs.harvard.edu/abs/2011MNRAS.418.1587T} {418, 1587}

\bibitem[\protect\citeauthoryear{{Theis}, {Burkert}  \& {Hensler}}{{Theis}
  et~al.}{1992}]{TBH92}
{Theis} C.,  {Burkert} A.,   {Hensler} G.,  1992, \aap, \href
  {http://esoads.eso.org/abs/1992A%26A...265..465T} {265, 465}

\bibitem[\protect\citeauthoryear{{Thornton}, {Gaudlitz}, {Janka}  \&
  {Steinmetz}}{{Thornton} et~al.}{1998}]{TGJ98}
{Thornton} K.,  {Gaudlitz} M.,  {Janka} H.~T.,   {Steinmetz} M.,  1998, \mn@doi
  [\apj] {10.1086/305704}, \href
  {https://ui.adsabs.harvard.edu/abs/1998ApJ...500...95T} {500, 95}

\bibitem[\protect\citeauthoryear{{Travaglio}, {Hillebrandt}, {Reinecke}  \&
  {Thielemann}}{{Travaglio} et~al.}{2004}]{Trav04}
{Travaglio} C.,  {Hillebrandt} W.,  {Reinecke} M.,   {Thielemann} F.~K.,  2004,
  \mn@doi [\aap] {10.1051/0004-6361:20041108}, \href
  {https://ui.adsabs.harvard.edu/abs/2004A&A...425.1029T} {425, 1029}

\bibitem[\protect\citeauthoryear{Tsujimoto}{Tsujimoto}{2011}]{Tsu11}
Tsujimoto T.,  2011, \mn@doi [The Astrophysical Journal]
  {10.1088/0004-637x/736/2/113}, 736, 113

\bibitem[\protect\citeauthoryear{{Tsujimoto} \& {Bekki}}{{Tsujimoto} \&
  {Bekki}}{2011}]{TsuB11}
{Tsujimoto} T.,  {Bekki} K.,  2011, \mn@doi [\aap]
  {10.1051/0004-6361/201016210}, \href
  {https://ui.adsabs.harvard.edu/abs/2011A&A...530A..78T} {530, A78}

\bibitem[\protect\citeauthoryear{{V{\"a}is{\"a}nen} et~al.,}{{V{\"a}is{\"a}nen}
  et~al.}{2008}]{VMK08}
{V{\"a}is{\"a}nen} P.,  et~al., 2008, \mn@doi [\mnras]
  {10.1111/j.1365-2966.2007.12703.x}, \href
  {http://esoads.eso.org/abs/2008MNRAS.384..886V} {384, 886}

\bibitem[\protect\citeauthoryear{{Valcke}, {de Rijcke}  \& {Dejonghe}}{{Valcke}
  et~al.}{2008}]{VDD08}
{Valcke} S.,  {de Rijcke} S.,   {Dejonghe} H.,  2008, \mn@doi [\mnras]
  {10.1111/j.1365-2966.2008.13654.x}, \href
  {http://esoads.eso.org/abs/2008MNRAS.389.1111V} {389, 1111}

\bibitem[\protect\citeauthoryear{{Vorobyov}, {Recchi}  \& {Hensler}}{{Vorobyov}
  et~al.}{2012}]{VRH12}
{Vorobyov} E.~I.,  {Recchi} S.,   {Hensler} G.,  2012, \mn@doi [\aap]
  {10.1051/0004-6361/201219113}, \href
  {http://esoads.eso.org/abs/2012A%26A...543A.129V} {543, A129}

\bibitem[\protect\citeauthoryear{{Vorobyov}, {Recchi}  \& {Hensler}}{{Vorobyov}
  et~al.}{2015}]{VRH15}
{Vorobyov} E.~I.,  {Recchi} S.,   {Hensler} G.,  2015, \mn@doi [\aap]
  {10.1051/0004-6361/201425587}, \href
  {http://esoads.eso.org/abs/2015A%26A...579A...9V} {579, A9}

\bibitem[\protect\citeauthoryear{{Webb} \& {Sills}}{{Webb} \&
  {Sills}}{2021}]{WS21}
{Webb} J.~J.,  {Sills} A.,  2021, \mn@doi [\mnras] {10.1093/mnras/staa3832},
  \href {https://ui.adsabs.harvard.edu/abs/2021MNRAS.501.1933W} {501, 1933}

\bibitem[\protect\citeauthoryear{{Weidner} \& {Kroupa}}{{Weidner} \&
  {Kroupa}}{2005}]{WK05}
{Weidner} C.,  {Kroupa} P.,  2005, \mn@doi [\apj] {10.1086/429867}, \href
  {http://esoads.eso.org/abs/2005ApJ...625..754W} {625, 754}

\bibitem[\protect\citeauthoryear{{Weidner} \& {Kroupa}}{{Weidner} \&
  {Kroupa}}{2006}]{WK06}
{Weidner} C.,  {Kroupa} P.,  2006, \mn@doi [\mnras]
  {10.1111/j.1365-2966.2005.09824.x}, \href
  {http://esoads.eso.org/abs/2006MNRAS.365.1333W} {365, 1333}

\bibitem[\protect\citeauthoryear{{Weidner}, {Kroupa}  \& {Bonnell}}{{Weidner}
  et~al.}{2010}]{WKB10}
{Weidner} C.,  {Kroupa} P.,   {Bonnell} I.~A.~D.,  2010, \mn@doi [\mnras]
  {10.1111/j.1365-2966.2009.15633.x}, \href
  {https://ui.adsabs.harvard.edu/abs/2010MNRAS.401..275W} {401, 275}

\bibitem[\protect\citeauthoryear{{Weidner}, {Kroupa}, {Pflamm-Altenburg}  \&
  {Vazdekis}}{{Weidner} et~al.}{2013}]{WKP13}
{Weidner} C.,  {Kroupa} P.,  {Pflamm-Altenburg} J.,   {Vazdekis} A.,  2013,
  \mn@doi [\mnras] {10.1093/mnras/stt1806}, \href
  {http://esoads.eso.org/abs/2013MNRAS.436.3309W} {436, 3309}

\bibitem[\protect\citeauthoryear{{Weisz} et~al.,}{{Weisz}
  et~al.}{2012}]{Weisz12}
{Weisz} D.~R.,  et~al., 2012, \mn@doi [\apj] {10.1088/0004-637X/744/1/44},
  \href {https://ui.adsabs.harvard.edu/abs/2012ApJ...744...44W} {744, 44}

\bibitem[\protect\citeauthoryear{{Weisz} et~al.,}{{Weisz}
  et~al.}{2015}]{Weisz15}
{Weisz} D.~R.,  et~al., 2015, \mn@doi [\apj] {10.1088/0004-637X/806/2/198},
  \href {https://ui.adsabs.harvard.edu/abs/2015ApJ...806..198W} {806, 198}

\bibitem[\protect\citeauthoryear{{Yagi}, {Gu}, {Fujita}, {Nakazawa}, {Akahori},
  {Hattori}, {Yoshida}  \& {Makishima}}{{Yagi} et~al.}{2013}]{YGF13}
{Yagi} M.,  {Gu} L.,  {Fujita} Y.,  {Nakazawa} K.,  {Akahori} T.,  {Hattori}
  T.,  {Yoshida} M.,   {Makishima} K.,  2013, ApJ, 778, 91

\bibitem[\protect\citeauthoryear{{de Grijs} \& {Anders}}{{de Grijs} \&
  {Anders}}{2006}]{deGA06}
{de Grijs} R.,  {Anders} P.,  2006, \mn@doi [\mnras]
  {10.1111/j.1365-2966.2005.09856.x}, \href
  {https://ui.adsabs.harvard.edu/abs/2006MNRAS.366..295D} {366, 295}

\bibitem[\protect\citeauthoryear{{de Grijs}, {Parmentier}  \& {Lamers}}{{de
  Grijs} et~al.}{2005}]{deGPL05}
{de Grijs} R.,  {Parmentier} G.,   {Lamers} H.~J.~G.~L.~M.,  2005, \mn@doi
  [\mnras] {10.1111/j.1365-2966.2005.09640.x}, \href
  {https://ui.adsabs.harvard.edu/abs/2005MNRAS.364.1054D} {364, 1054}

\makeatother
\end{thebibliography}

\bsp	
\label{lastpage}
\end{document}